\documentclass[fleqn,usenatbib]{mnras} 

\usepackage{newtxtext,newtxmath,aas_macros}
\usepackage[para,online,flushleft]{threeparttable}

\usepackage[T1]{fontenc}
\usepackage{ae,aecompl}

\usepackage{graphicx}	
\usepackage{amsmath}	
\usepackage{amssymb}	

\usepackage{xfrac}	
\usepackage{tablefootnote}
\usepackage{booktabs,caption}

\def\mstar{$m^*$}

\def\Rtwo    {$r_{200}$}
\def\Rfive   {$r_{500}$}
\def\BCGd {D$_{\rm BCG-SZ}$}
\def\BCGXd {D$_{\rm BCG-X}$}
\def\Aphot {A$_{\rm phot}$}

\def\allpop{global}
\def\allpop{all}
\def\sqd{deg$^2$}

\def\Nclusters {368}
\def\NclustersMZcut{288}
\def\NclustersMZcutNoUnrelaxedNorelaxed{204} 

\def\DESXVPNurgalievlimited{41}

\def\ChandraTeslaNEW{1} 

\def\NChandralimited {42} 

\def\ChandraXMMNtotalYthreelimited{67}

\def\SPTLinPvalue{0.094} 
\def\SPTLinDvalue{0.072} 
\def\Ncoolcore {41}
\def\Ndisturbed {43} %
\def\NdisturbedSZonly {39} 

\def\NdisturbedHighZ {17} %
\def\NdisturbedLowZ {26} %
\def\NcoolcoreHighZ {25}
\def\NcoolcoreLowZ {16}
\def\NAllexclHighZ {93} %
\def\NAllexclLowZ {111} %

\def\NAphotDISTlimited{7} 

\def\NAphotRELAXlimited{15}

\def\MDcoolcoreLim{16}
\def\MDcoolcoreLimInAPHOT{9}

\def\TeslaCHANDRAlimited{39}

\def\NXMMDISTcentroidlimitedNEW{1}

\def\NXMMCCcentroidlimitedNEW{12}

\def\DESXMMN{43}
\def\DESXMMNlimited{28}
\def\DESXMMNlimitedNEW{25}

\def\ChandraNEWpeakRelaxedNEW{11} 

\def\Nspec {110}
\def\NspecG{33}
\def\NspecNG{9}
\def\Nspectocompare{42} 
\def\NspecHere {104}
\def\BCGallMATCH{68} 
 
\def\RMSPTMATCHES{245}
\def\RMSPTMATCHESPct{85} 
\def\RMSPTMISSING{43} 
\def\RMSPTMISSINGvoids{24} 
\def\RMSPTMATCHESPctII{93} 
\def\RMBCGMATCHES{168} 
\def\RMBCGMATCHESmiss{78}

\def\RMBCGMATCHESmissCentralBCG{25}
\def\RMBCGright{2} 

\def\RMBCGright{1} 
\def\RMBCGundef{$7$} 
\def\RMBCGMATCHESmissFainterBCG{45} 

\def\KSMergingIntermediate{0.2736}
\def\KSRelaxedIntermediate{0.0083}
\def\KSMergingRelaxed{0.0028}

\def\KSRelaxedIntermediateLowz{0.64}
\def\KSMergingRelaxedLowz{0.32}
\def\RMBCGMATCHESmissTWO{70}  
\def\RMBCGMATCHESmissTWOPct{72} 

\def\DESdepthgTen{23.45}
\def\DESdepthrTen{23.10}
\def\DESdepthiTen{22.44}
\def\DESdepthzTen{21.71}

\def\DESdepthgTen{$23.62 \pm 0.16$}
\def\DESdepthrTen{$23.34 \pm 0.15$}
\def\DESdepthiTen{$22.78 \pm 0.16$}
\def\DESdepthzTen{$22.13 \pm 0.16$}

\def\DESdepthgFive{$24.37 \pm 0.16$}
\def\DESdepthrFive{$24.11 \pm 0.14$}
\def\DESdepthiFive{$23.58 \pm 0.16$}
\def\DESdepthzFive{$22.92 \pm 0.16$}

\def\HONNORMLINPIVOTALLCLSthreeSigmaALL{$32.4^{+1.5}_{-2.2}$}
\def\HONNORMLINPIVOTunrelaxedALLthreeSigma{$32.4^{+5.7}_{-4.8}$}
\def\HONNORMLINPIVOTrelaxedALLthreeSigma{$31.6 \pm 4.7 $}
\def\HONNORMLINPIVOTALLCLSREDthreeSigmaALL{$24.0^{+1.7}_{-1.6}$}
\def\HONNORMLINPIVOTunrelaxedREDthreeSigmaALL{$23.4^{+4.7}_{-3.1}$}
\def\HONNORMLINPIVOTrelaxedREDthreeSigmaALL{$24.0^{+4.2}_{-3.6}$}
\def\HONNORMLINPIVOTALLCLSthreeSigmaHalf{$17.0^{+1.6}_{-1.1}$}

\def\HONNORMLINPIVOTunrelaxedALLthreeSigmahalf{$16.2^{+5.2}_{-4.2}$}
\def\HONNORMLINPIVOTrelaxedALLthreeSigmahalf{$17.4^{+3.5}_{-2.9}$}
\def\HONNORMLINPIVOTREDCLSthreeSigmaHalf{$13.5^{+1.7}_{-1.2}$}
\def\HONNORMLINPIVOTunrelaxedREDthreeSigmahalf{$12.3^{+5.5}_{-3.6}$}
\def\HONNORMLINPIVOTrelaxedREDthreeSigmahalf{$13.5^{+3.5}_{-2.8}$}

\def\BlueDisturbedBCGs{3} 
\def\BlueDisturbedBCGsPct{7}

\def\BlueTOTALBCGsPct{2}
\def\BlueTOTALBCGsPIPINORANGEPct{1}

\def\SigmaZeroRtwo{$0.05\pm 0.01$}
\def\SigmaOneRtwo{$0.35\pm 0.03$}

\def\SigmaZeroRfive{$0.08\pm0.01$}
\def\SigmaOneRfive{$0.55\pm0.04$}
\usepackage{eso-pic}

\AddToShipoutPictureBG*{%
  \AtPageUpperLeft{%
    \hspace{0.75\paperwidth}%
    \raisebox{-3.5\baselineskip}{%
      \makebox[0pt][l]{\textnormal{DES-2017-0323}}
}}}%

\AddToShipoutPictureBG*{%
  \AtPageUpperLeft{%
    \hspace{0.75\paperwidth}%
    \raisebox{-4.5\baselineskip}{%
      \makebox[0pt][l]{\textnormal{FERMILAB-PUB-20-097-AE}}
}}}%

\title[Joint SZ-Xray-optical analysis of \NclustersMZcut\ SPT clusters]{A joint SZ-Xray-optical analysis of the dynamical state of \NclustersMZcut\ massive galaxy clusters}

\author[DES Collaboration]{
\parbox{\textwidth}{
\Large
A.~Zenteno,$^{1}$
D.~Hern\'andez-Lang,$^{2,1,3}$
M.~Klein,$^{2,4}$
C.~Vergara Cervantes,$^{5}$
D.~L.~Hollowood,$^{6}$
S.~Bhargava,$^{5}$
A.~Palmese,$^{7,8}$
V.~Strazzullo,$^{2,9}$
A.~K.~Romer,$^{5}$
J.~J.~Mohr,$^{2,4}$
T.~Jeltema,$^{6}$
A.~Saro,$^{9,10,11}$
C.~Lidman,$^{12}$
D.~Gruen,$^{13,14,15}$
V.~Ojeda,$^{16}$
A.~Katzenberger,$^{17}$
M.~Aguena,$^{18,19}$
S.~Allam,$^{7}$
S.~Avila,$^{20}$
M.~Bayliss,$^{21}$
E.~Bertin,$^{22,23}$
D.~Brooks,$^{24}$
E.~Buckley-Geer,$^{7}$
D.~L.~Burke,$^{14,15}$
A.~Carnero~Rosell,$^{25}$
M.~Carrasco~Kind,$^{26,27}$
J.~Carretero,$^{28}$
F.~J.~Castander,$^{29,30}$
M.~Costanzi,$^{11,31}$
L.~N.~da Costa,$^{19,32}$
J.~De~Vicente,$^{25}$
S.~Desai,$^{33}$
H.~T.~Diehl,$^{7}$
P.~Doel,$^{24}$
T.~F.~Eifler,$^{34,35}$
A.~E.~Evrard,$^{36,37}$
B.~Flaugher,$^{7}$
B.~Floyd,$^{38}$
P.~Fosalba,$^{29,30}$
J.~Frieman,$^{7,8}$
J.~Garc\'ia-Bellido,$^{20}$
D.~W.~Gerdes,$^{36,37}$
J.R.~Gonzalez,$^{39}$
R.~A.~Gruendl,$^{26,27}$
J.~Gschwend,$^{19,32}$
G.~Gutierrez,$^{7}$
W.~G.~Hartley,$^{40,24,41}$
S.~R.~Hinton,$^{42}$
K.~Honscheid,$^{43,44}$
D.~J.~James,$^{45}$
K.~Kuehn,$^{46,47}$
O.~Lahav,$^{24}$
M.~Lima,$^{18,19}$
M.~McDonald,$^{48}$
M.~A.~G.~Maia,$^{19,32}$
M.~March,$^{49}$
P.~Melchior,$^{50}$
F.~Menanteau,$^{26,27}$
R.~Miquel,$^{51,28}$
R.~L.~C.~Ogando,$^{19,32}$
F.~Paz-Chinch\'{o}n,$^{52,27}$
A.~A.~Plazas,$^{50}$
A.~Roodman,$^{14,15}$
E.~S.~Rykoff,$^{14,15}$
E.~Sanchez,$^{25}$
V.~Scarpine,$^{7}$
M.~Schubnell,$^{37}$
S.~Serrano,$^{29,30}$
I.~Sevilla-Noarbe,$^{25}$
M.~Smith,$^{53}$
M.~Soares-Santos,$^{54}$
E.~Suchyta,$^{55}$
M.~E.~C.~Swanson,$^{27}$
G.~Tarle,$^{37}$
D.~Thomas,$^{56}$
T.~N.~Varga,$^{4,2}$
A.~R.~Walker,$^{1}$
and R.D.~Wilkinson$^{5}$
\begin{center} (DES Collaboration) \end{center}
}
\vspace{0.4cm}
\\
\parbox{\textwidth}{
\scriptsize
$^{1}$ Cerro Tololo Inter-American Observatory, NSF's National Optical-Infrared Astronomy Research Laboratory, Casilla 603, La Serena, Chile\\
$^{2}$ Faculty of Physics, Ludwig-Maximilians-Universit\"at, Scheinerstr. 1, 81679 Munich, Germany\\
$^{3}$ Gemini Observatory, NSF's National Optical-Infrared Astronomy Research Laboratory, Casilla 603, La Serena, Chile\\
$^{4}$ Max Planck Institute for Extraterrestrial Physics, Giessenbachstrasse, 85748 Garching, Germany\\
$^{5}$ Department of Physics and Astronomy, Pevensey Building, University of Sussex, Brighton, BN1 9QH, UK\\
$^{6}$ Santa Cruz Institute for Particle Physics, Santa Cruz, CA 95064, USA\\
$^{7}$ Fermi National Accelerator Laboratory, P. O. Box 500, Batavia, IL 60510, USA\\
$^{8}$ Kavli Institute for Cosmological Physics, University of Chicago, Chicago, IL 60637, USA\\
$^{9}$ Astronomy Unit, Department of Physics, University of Trieste, via Tiepolo 11, I-34131 Trieste, Italy\\
$^{10}$ IFPU - Institute for Fundamental Physics of the Universe, Via Beirut 2, 34014 Trieste, Italy\\
$^{11}$ INAF-Osservatorio Astronomico di Trieste, via G. B. Tiepolo 11, I-34143 Trieste, Italy\\
$^{12}$ The Research School of Astronomy and Astrophysics, Australian National University, ACT 2601, Australia\\
$^{13}$ Department of Physics, Stanford University, 382 Via Pueblo Mall, Stanford, CA 94305, USA\\
$^{14}$ Kavli Institute for Particle Astrophysics \& Cosmology, P. O. Box 2450, Stanford University, Stanford, CA 94305, USA\\
$^{15}$ SLAC National Accelerator Laboratory, Menlo Park, CA 94025, USA\\
$^{16}$ Departamento de F\'isica, Universidad de La Serena, La Serena, Chile\\
$^{17}$ Department of Geography, Ludwig-Maximilians-Universit\"at, Luisenstr. 37, 80333 Munich, Germany\\
{\it Continues at the end of the paper.}
}
}

\date{Accepted 2020 April 22. Received 2020 April 21; in original form 2020 April 3}

\pubyear{2020}

\begin{document}
\label{firstpage}
\pagerange{\pageref{firstpage}--\pageref{lastpage}}
\maketitle

\begin{abstract}
We use imaging from the first three years of the Dark Energy Survey  to characterize the dynamical state of \NclustersMZcut\ galaxy clusters at  $0.1 \lesssim z \lesssim 0.9$ detected in the South Pole Telescope  (SPT) Sunyaev-Zeldovich (SZ) effect survey (SPT-SZ). We examine spatial offsets between the position of the brightest cluster galaxy  (BCG) and the center of the gas distribution as traced by the SPT-SZ centroid and by the X-ray centroid/peak position from Chandra and XMM data.  We show that the radial distribution of offsets provides no evidence that SPT SZ-selected cluster samples include a higher fraction of  mergers  than X-ray-selected cluster samples. We use the offsets to classify the dynamical state of the clusters, selecting the \Ndisturbed\ most disturbed clusters, with half of those at $z \gtrsim 0.5$, a region seldom explored previously.  We find that Schechter function fits to the galaxy population in disturbed clusters and relaxed clusters differ at $z>0.55$ but not at lower redshifts. Disturbed clusters at $z>0.55$ have steeper faint-end slopes and brighter characteristic magnitudes.  Within the same redshift range, we find that the BCGs in relaxed clusters tend to be brighter than the BCGs in disturbed samples, while in agreement in the lower redshift bin. Possible explanations includes a higher merger rate, and a more efficient dynamical friction at high redshift. The red-sequence population is less affected by the cluster dynamical state than the general galaxy population.
\end{abstract}


\section{Introduction}
Galaxy cluster mergers are powerful events. The energy released can be up to the order of $\gtrsim 10^{64}$ ergs \citep{sarazin02}, and the  effects of the event can be observed from radio wavelengths, in the form of radio haloes and relics \citep[e.g.,][]{ensslin98,cassano10,vanweeren11,drabent15,eckert17}, to X--ray wavelengths 
\citep[e.g.,][]{markevitch07,takizawa10,nelson14}.  
Cluster mergers provide unique conditions  to study a range of physics, from particle physics, to galaxy evolution and cosmology. They may play a role in accelerating cosmic rays \citep[e.g.,][]{vanweeren17}, generating gamma rays \citep[e.g.,][]{blasi07,pinzke10}, and can be used to constrain  the dark matter particle self-interaction cross-section 
\cite[e.g.,][]{markevitch04,harvey15,monteiro18}.  From a cosmological perspective, disturbed clusters  provide tests for $\Lambda$CDM \citep{thompson15,kimstacy17}, and they open a window into the past thanks to their enhanced strong lensing efficiency \citep{zitrin13,acebron19}.

From the perspective of galaxy evolution,  disturbed clusters provide  unique environmental conditions under which galaxies are transformed. The  abundance of jellyfish galaxies in disturbed clusters \citep{owers12,mcpartland16} shows that they can be used to examine the effects of the  ram pressure from the cluster ICM.  Although  jellyfish galaxies represent the most spectacular cases, extreme ram pressure may also have an impact on the overall cluster galaxy population.
While studies of the cluster luminosity function (LF) abound, studies of the changes in the galaxy population properties as a function of the cluster dynamical state are more scarce. They have been explored in single clusters \cite[e.g.,][]{ma10,pranger13,pranger14}, in samples in the low-mass, low-redshift regime \citep{ribeiro13,wen15}, and in small samples with a higher mass range and wider redshift range \citep{barrena12,depropris13}. Here  we further improve on such studies by extracting the most extreme cases from a much larger sample, and exploring the highest mass end of the cluster mass function \citep{bleem15a,bayliss16}, within a wide redshift range. 

Classification of disturbed systems can be done using several proxies: 
for example, the shape of the velocity dispersion  \citep[Gaussian versus non-Gaussian; e.g.,][]{martinez12,ribeiro13,delosrios16}, the X--ray morphology \citep[e.g.,][]{mohr93,jeltema05,bohringer10,rasia12,nurgaliev13,parekh15,nurgaliev17}, diffuse radio emission \citep[e.g.,][]{cassano10,feretti12}, the cluster galaxy density distribution \citep[e.g.,][]{zitrin11,wen15}, and a combination of observations at different wavelengths such as the X--ray peak/centroid to BCG offset \citep[e.g.,][]{mann12}.
%
The use of the BCG position as a proxy for the collisionless component is a logical choice (and less expensive in term of telescope time), as it is expected that the BCG will rapidly sink to the bottom of the potential well due to dynamical friction \citep{tremaine90}.  This is supported by several observational studies, either when the center is measured by weak lensing analysis \citep[e.g.,][]{oguri10,zitrin12}, or by gas based measurements \citep{lin04b,hudson10,song12b,mann12,rozo14}.  Such studies have shown that the majority of the BCGs lie close to the collisional component (the gas; within  50 $h^{-1}$ kpc), and that the remaining clusters consist of dynamically disturbed systems, with large offsets.  Furthermore, 
hydrodynamical cosmological simulations have also shown that the BCG to dark matter offset is the tightest dynamical state proxy among several studied \citep{ng17,cialone18}. In this work, we thus adopt the BCG as the center of the collisionless component.

Large cluster samples are typically identified using optical,  X--ray, and Sunyaev-Zeldovich (SZ) surveys. 
Currently, large SZ samples of galaxy clusters are  available;  the {\it Planck} telescope \citep{planck16-26}, the Atacama Cosmology Telescope \citep{hasselfield13}, and the South Pole Telescope  \citep[SPT;][]{bleem15a} altogether have detected about 2000 galaxy clusters. 
In this work, we use the SPT cluster sample, and the SZ centroid as a proxy for the collisional component. 

The positions of the BCGs are provided by analyzing data from the first three years of the  Dark Energy Survey \citep[DES;][]{des16,abbott18,morganson18}. DES is an imaging survey designed to provide constraints on cosmological parameters using four cosmological probes: galaxy clusters, baryon acoustic oscillation, weak lensing, and type Ia supernovae.  To  achieve this goal, the final depth (which will come after Year 6) is estimated to be $\sim$ 24 AB mag in the $grizY$ bands over a continuous 5,000 \sqd\ area in the southern sky including the SPT-SZ survey footprint.  This dataset is ideal for finding  BCGs within the cluster virial radius, and systematically characterizing the cluster galaxy properties and the local background of massive clusters up to redshift $\sim$ 1.

This paper is organized as follows. In \S~\ref{sec:clustersample} we introduce the data and the cluster sample. In \S~\ref{sec:proxies} we describe in detail the dynamical state proxies used. In \S~\ref{sec:galpop} we describe the tools we use to statistically characterize the galaxy population, while in  \S~\ref{sec:results} we show our results.  In  \S~\ref{sec:conclusions} we present our conclusions.

Throughout this work, we assume a flat $\Lambda$CDM cosmology with $H_0$=68.3 km s$^{-1}$ Mpc$^{-1}$ and $\Omega_{\rm M}$ = 0.299 \citep{bocquet15}.

\section{Data and measurements}
\label{sec:clustersample}

\subsection{SPT-SZ cluster sample}
\label{sec:spt} 
The cluster sample comes from the 2,500 deg$^2$ SPT-SZ survey. It consists of 516 (387) optically-confirmed clusters at signal-to-noise $> 4.5 (5)$ \citep[][hereafter B15]{bleem15a}. The SPT-SZ cluster sample can be  considered as approximately  mass selected with a nearly redshift independent mass threshold of M$_{\rm 200}$ $> 4 \times 10^{14}$ M$_\odot$. The redshift range extends to $z\approx$1.7 \citep{khullar19,strazzullo19}. We adopt as an estimate of cluster mass M$_{\rm 200,critical}$, which is the mass within a clustercentric distance of \Rtwo, with \Rtwo\ being the radius within which the mean density is 200 times the critical density of the Universe at the cluster redshift. We estimate M$_{\rm 200,critical}$ from the SZE-based M$_{\rm 500,critical}$ published in \cite{bleem15a}, adopting the \cite{duffy08} mass-concentration relation.

\subsection{DECam optical imaging}
\label{sec:desimaging}
Here we use Dark Energy Camera \citep*[DECam;][]{flaugher15} images from the  DES Y3 data release, taken  from August 31, 2013 to February 12, 2016 \citep[henceforth DESY3;][]{abbott18}, covering $\sim5,000$ \sqd. The data are processed through the DES Data Management (DESDM) system \citep{morganson18}, which detrends and calibrates the raw DES images, combines individual exposures to create coadded images, and detects and performs photometry of astrophysical objects. 

The DES data are key for this study, as they allow us to search for the BCG within the  \Rtwo\  for clusters in a wide redshift range.  Such a wide redshift range translates into a very different projected radius on the sky.  The DES  data provides us with complete coverage of \Rtwo\ and a robust estimate of the local galaxy background for each cluster.

We cross match the SPT-SZ catalog (at signal to noise $> 4.5$) with the DESY3 Gold sample catalog (Sevilla-Noarbe, in preparation), excluding clusters with missing information (large gaps, missing bands, bright foreground stars close to the cluster center, etc.) or bad photometry, finding \Nclusters\ galaxy clusters.  The mean 10(5)$\sigma$ depth of the multi-object PSF-fitting photometry \citep{drlica18}, and the scatter (1$\sigma$)  across the cluster fields, are \DESdepthgTen\ (\DESdepthgFive), \DESdepthrTen\ (\DESdepthrFive), \DESdepthiTen\ (\DESdepthiFive), and \DESdepthzTen\ (\DESdepthzFive) for $griz$, respectively,  providing an average depth of $m^*+2$ up to redshift $0.66 \pm 0.04$ ($0.83 \pm 0.06$).

\subsection{Redshifts and cluster masses}
\label{sub:zmass}
\begin{figure}
	\includegraphics[scale=0.58]{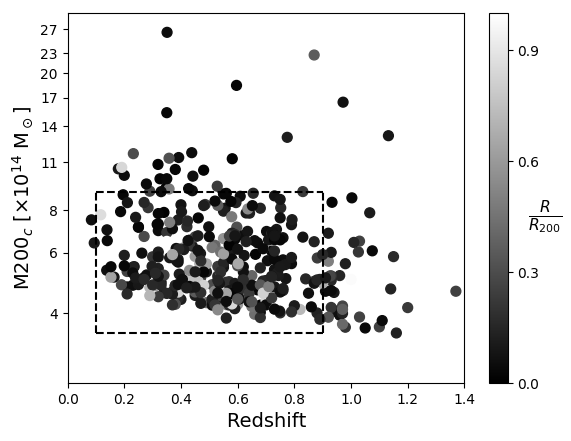}
    \caption{The mass redshift distribution of  \Nclusters\ SPT clusters imaged with DES Y3 data.  The dashed box corresponds to limits applied to the SPT-SZ sample to select our sample.  In light gray clusters with larger BCG to SZ centroid  offset (\BCGd) while the dark points correspond to systems with lower \BCGd. }
    \label{fig:distribution}
\end{figure}

Clusters redshifts, photometric and spectroscopic, are collected from several sources including \cite{planck11-9}, \cite{sifon13}, \cite{bleem15a}, and \cite{bayliss16}. In total there are \Nspec\ SPT clusters with spectroscopic redshifts in the SPT-SZ sample \citep{capasso19}.  The  cross match of the \Nclusters\ optically cross-matched clusters with the spectroscopic sample renders \NspecHere\ clusters.  The masses adopted here are drawn from B15 and transformed to M$_{200}$, as described in \S~\ref{sec:spt}, and the mass and redshifts distribution of the \Nclusters\ clusters is shown in Fig~\ref{fig:distribution}. 
  
As we explore the cluster population as a function of dynamical state, we introduce mass-redshift cuts.
%
As we shall see in \S~\ref{sec:proxies}, there is a clear tendency for relaxed clusters to be more  massive.  This is because we selected relaxed clusters using an X--ray proxy, and most of the X--ray data come from a study designed to investigate the most massive systems in SPT \citep{mcdonald13}. These cuts are applied to all samples discussed in this paper. 
The redshift and mass cuts, designed to limit mass and/or redshift dependencies when comparing sub samples, are z $\in$ [0.1:0.9] and M$_{200} < 9\times 10^{14}$ M$_{\odot}$.  From works such as \cite{hennig17} and \cite{martinet17}, we expect that the impact on the luminosity function from the selection, in that redshift-mass range, will be negligible. These cuts render a final number of \NclustersMZcut\ clusters, hereafter referred to as the DESY3-SPT sample. 

\subsection{BCG selection}
\label{sec:bcgselection}
%
BCG selection was made in two  ways: automatically and by visual inspection of DES images\footnote{We use pseudo-color images using $gri$ for clusters at $z<0.35$ and $riz$ for systems above that redshift.}. The latter method involves visually selecting all the potential cluster galaxies within \Rtwo, based on several properties including: size, colors, and the number of neighbouring galaxies. We then select the brightest galaxy as the BCG.  The automatic method instead proceeds by first fitting a passively evolving simple stellar population (SSP) model, as adopted in \citet[][hereafter Z16]{zenteno16} and in \citet[][hereafter H17]{hennig17}\footnote{We use the \cite{bruzual03} synthesis models, assuming a single burst of star formation at z = 3 followed by passive evolution to z = 0, with six distinct metallicities to match the tilt of the color-magnitude relation at low redshift. This model  correctly reproduces the observed redshift evolution of \mstar\ in the probed redshift range.}, to the red-sequence (RS) and then selecting the brightest RS galaxy as the BCG, within \Rtwo. More specifically, we first fit the red sequence model\footnote{ We use $g-r$ for $z<0.30$, $r-i$ for $0.3 \leq  z \leq 0.7$, and $i-z$ for $z > 0.7$} to all galaxies within $0.25\times$\Rtwo\ of the SZ center and select galaxies with colors that lie within a generous\footnote{This corresponds to a 3$\sigma$ average RS scatter in magnitudes \citep{lopez04}, or 3$\sigma$ intrinsic RS  scatter observed in DES data, on SPT clusters, at z $\approx 1$  \citep{hennig17}.} margin of $\pm 0.22$ of the red sequence fit.  To find the best RS fit to our model we use the B15 redshifts as a prior with a delta of $\pm$ 0.15. The redshift estimation method employed in B15 is essentially the same as ours, and we do not expect significant differences. The delta of 0.15 is thus a generous margin, a few times the typical difference.  Once the best fitting model is found, we  use this RS model to select RS galaxies, within $\pm$ 0.22 magnitudes in color within \Rtwo, identifying the brightest red galaxy among them.  
We limit the search at the bright end to $m^*-4$ to limit foreground contamination. Our experience with SPT clusters has not shown BCGs brighter than that. With these two lists of BCGs, we then inspect the discrepancies to decide, case by case, if the automatic method chose the wrong BCG (e.g., a foreground source with a color consistent with the red sequence), or if the visual selection missed a cluster member.  This is an iterative process that has involved three of us comparing and checking BCGs, and across several DES data releases and photometric versions.  The final BCGs are chosen after this process. 

We  compare our selection to the RedMaPPer Y3 (RMY3) BCG selection.  We match our DESY3-SPT cluster sample to RMY3 using a  \Rtwo\  search radius and a richness ($\lambda$) $>$ 20 cut, finding \RMSPTMATCHES\ matches (\RMSPTMATCHESPct\%).   
Out of the \RMSPTMISSING\ missing systems, \RMSPTMISSINGvoids\ are in extended  areas masked by RM.  If we do not consider these, the percentage of DESY3-SPT clusters identified by RM rises to \RMSPTMATCHESPctII\%, with most of the other non-matches being lower mass clusters under different configurations (close to stars or big galaxies, at $z > 0.6$, and even one including a cluster pair in which RM detects just one). 
%
From that sample we match the  RMY3 BCGs, finding \RMBCGMATCHES\ RMY3-SPT BCG matches  and \RMBCGMATCHESmiss\ mismatches. We then proceed to perform a visual and catalog inspection of the \RMBCGMATCHESmiss\ mismatches to evaluate potential errors. 
Of the \RMBCGMATCHESmiss\ mismatches, \RMBCGMATCHESmissCentralBCG\ correspond to the brightest galaxy within 0.2\Rtwo\ from the SPT-SZ centroid, but fainter than our selection within \Rtwo, while \RMBCGMATCHESmissFainterBCG\ are just fainter than our BCG candidate.  Only \RMBCGright\ RMY3 BCG seems to be undoubtedly a better choice. Our candidate BCG is brighter at the catalog level, but a visual inspection reveals that this is clearly a mistake.  RM used single epoch photometry and it seems that in this case it performed better than the stacked photometry for this particular BCG.  Furthermore, there are \RMBCGundef\ cases in which the BCG selection between RM and ours is different, but they are so similar visually and at catalog level, that the difference could be accounted for by the different photometric catalogs used.  If we discard those cases, we have \RMBCGMATCHESmissTWO\ mismatches, rendering a  matching success between our BCG selection and the RM-BCG of  \RMBCGMATCHESmissTWOPct\%. This is consistent with the  results of \cite{hoshino15}, which finds that 20\% to 30\% of the RedMaPPer central galaxies are not the brightest cluster member.

We also estimate the level of contamination of our  BCG selection statistically.  We use the cluster RS to select galaxies brighter than our candidate BCG in a surrounding area equivalent to 8 times the cluster area (1-3 \Rtwo), filtering stars by using entries with the {\tt EXTENDED\_CLASS\_MASH\_SOF} parameter $\geq 2$ (which excludes high confidence stars and candidate stars). This scenario is conservative; no visual inspection was carried out to discard non-galaxy looking candidates and we would not expect a BCG being located that far from the cluster gas component. 
For \NclustersMZcut\ clusters we find zero brighter galaxies for 26\% ($p_0$) of the clusters, for 32\% ($p_1$) we find 1 galaxy brighter, for 23\% ($p_2$) we find 2 galaxies, for 10\% ($p_3$) we find  3 brighter galaxies, for 6\% ($p_4$) we find 4 brighter galaxies, for 2\% ($p_5$)  we find 5 brighter galaxies, while for 1\% ($p_6$)  we find 6 brighter galaxies.  There are no systems with more than 6 brighter galaxies in the background area used.  Adding the probabilities, using the number of red-sequence galaxies brighter than our chosen BCG ($N_{{\rm BCG},i}^{\rm back}$), and normalizing by the area, we find a $P_{\rm tot} =1 -   \sum_{i=0}^{6}  p_i \frac{1}{1 + N_{{\rm BCG},i}^{\rm back}/8} = 14$ \% chance of BCG miss-identification over the whole sample.

Finally, we test the accuracy of the method by cross matching our BCGs to  spectroscopic redshifts found in the literature as  described in \S~\ref{sub:zmass} \citep{bayliss14,bayliss16,capasso19}.  We match the BCG sample and the spectroscopic samples within 2 arcseconds, finding \BCGallMATCH\ candidate BCGs with redshifts.   From this match we find only one system with an inconsistent redshift;  SPT-CLJ2301-4023 has a BCG with a redshift of 0.778069 while the cluster redshift is 0.8349.  A visual inspection reveals that the BCG belongs to the red sequence and looks like a cluster member surrounded by several other cluster members with the same color.  The spectroscopic BCG is also red and bright with a magnitude consistent with the visual BCG within the photometric error bars.  Extrapolating to the whole sample, this implies that about 2\% of the BCGs selected with our procedure may be misidentified.  We further test our BCG selection in \S~\ref{sub:BCGtests}.

\subsection{X--ray data}
\label{sec:xray}

\subsubsection{{\it Chandra}}
\label{sec:chandra}

The {\it Chandra} X-ray Visionary Project (XVP; PI B. Benson) published X--ray data of 90 SPT systems, at $0.35 < z < 1.2$, and their properties have been characterized in several papers \citep[e.g, ][]{semler12,mcdonald14}, including coolcore  properties  \citep{mcdonald13} and a  morphological classification \citep{nurgaliev17}, which are relevant for this study.  
The matching of the \cite{nurgaliev17} sample with DESY3-SPT renders \DESXVPNurgalievlimited\ clusters within our redshift-mass limits (DESY3-XVP sample hereafter). 

We also make use of  Chandra  public data by using the \textbf{M}ass \textbf{A}nalysis \textbf{T}ool for \textbf{Cha}ndra (MATCha) pipeline \citep{hollowood19}. Given an input cluster catalog, this code queries the Chandra archive and reduces all public data for clusters in the catalog.  The output of MATCha includes measurements of cluster $T_X$, $L_X$, and X-ray centroid within a set of apertures, 500 kpc, $r_{2500}$, $r_{500}$, and core-cropped $r_{500}$.  The position of the X-ray peak is estimated as the brightest pixel after smoothing with a Gaussian of 50 kpc width \citep[see][for details]{hollowood19}. Matching the MATCha program results to our DESY3-SPT sample we obtain \TeslaCHANDRAlimited\ matches, \ChandraTeslaNEW\ of them a new addition to our sample.

In total, we find \NChandralimited\ systems with {\it Chandra} data in the DESY3-SPT sample.

\subsubsection{XMM Newton}
\label{sec:xmmn}
We also searched for counterparts in the  XMM-Newton (XMM) space telescope archive. The XMM analysis was performed using an adaptation of the X-ray Automated Pipeline Algorithm \citep[XAPA;][]{freeman02} developed for the XMM Cluster Survey \citep[XCS, e.g,][]{romer00}. XCS uses all available X-ray data in the XMM Science Archive to search  for galaxy clusters that were detected first in XMM images. X-ray sources are detected in XMM images using an algorithm based on wavelet transformations \citep[e.g.,][]{lloyddavies11}, and then compared to the position dependent point spread function (PSF) to classify them as extended, PSF-sized, or point like. Most of the extended X-ray sources are clusters, therefore all extended sources are flagged as cluster candidates, but in some cases these can be low-redshift galaxies, supernova remnants or multiple blended point sources. 
We then cross match the DESY3-SPT clusters and XCS cluster candidates. We made the following assumptions: all physical matches will occur within 1.5 Mpc of the SPT center (assuming that the X-ray source to be at the SPT redshift),  and all physical matches will be contained within the sub-set of XCS sources that are defined as being extended and with more than 100 photon counts. Not all of the DESY3-SPT clusters in the XMM footprint will have such counterparts in XCS because either: (i) the different cluster finding methodology employed by the two surveys, or (ii) the respective XMM observation has a low exposure time and/or high background, or (iii) the SPT cluster falls on the edge of the field of view and/or in an EPIC chip gap.  This procedure renders \DESXMMN\  matches.  From those \DESXMMN\ matches, \DESXMMNlimited\ are found in the DESY3-SPT sample; \DESXMMNlimitedNEW\ of them provide new X--ray information to add to the {\it Chandra} list.

The final list of clusters with X--ray information in the DESY3-SPT sample is \ChandraXMMNtotalYthreelimited.

\section{dynamical state proxies}
\label{sec:proxies}
As clusters grow, by merging with other clusters or groups, the gas, dark matter, and galaxy components depart from relaxation. Therefore, signatures of that departure  can be identified by using a gas proxy and a (dark) matter proxy. Here we  describe the combination of X--ray, SZ and optical imaging proxies used to classify the clusters' dynamical state.

\subsection{Disturbed sample}
\subsubsection{BCG-SZ centroid offset}
\label{subsec:bcgsz} 

\begin{figure}
 	\includegraphics[scale=0.63]{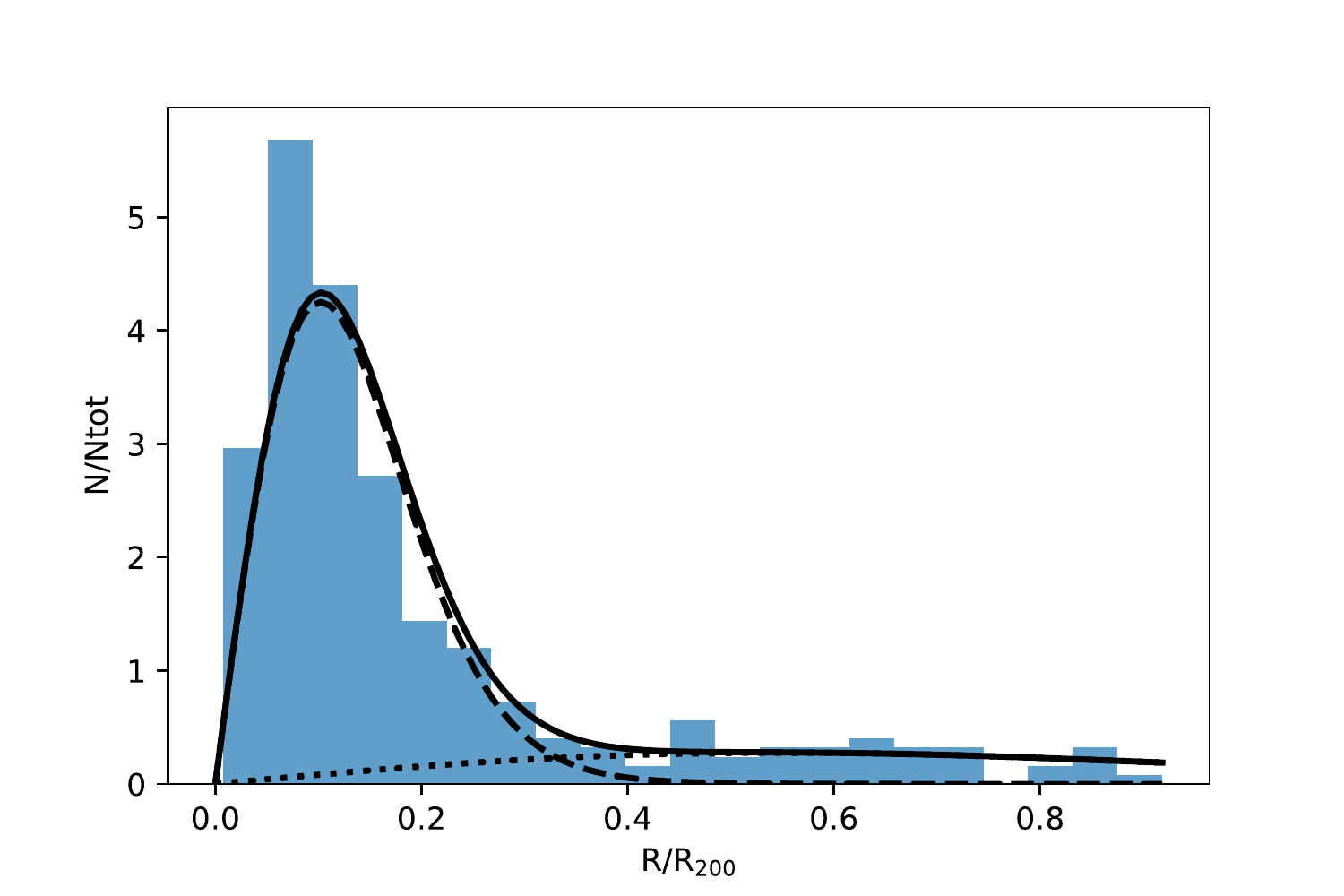}
    \caption{Normalized histogram of the BCG-SZ offset distribution. The \BCGd\ distribution can be described as the sum (black line) of a relaxed distribution (dashed line) plus a disturbed distribution (dotted line) The disturbed sample consists of the clusters with R~>~0.4~$\times$~\Rtwo\ and \BCGd\  > 3$\Delta\theta$.}
     \label{fig:arcsecdistribution}
\end{figure}

We use the offset between the BCG and the SPT-SZ centroid to classify clusters with the largest offsets as disturbed.  Specifically, we select all clusters that have a BCG-SZ offset of \BCGd\ > $0.4\times$\Rtwo.   The choice of $0.4\times$\Rtwo\ is taken after plotting the \BCGd\ histogram distribution, which looks flat after $\sim 0.4\times$\Rtwo\ (see Fig.~\ref{fig:arcsecdistribution}). We also take into account the SPT positional uncertainty $\Delta\theta$, which is defined by  $\sqrt{(\theta_{beam}^2+(k\theta_c)^2)}/\xi$, where $\xi$ is the significance of the cluster SZ detection, $\theta_{\rm beam}$ is the beam FWHM, $\theta_c$ is the core radius, and $k$ is a factor of the order of unity \citep[see][]{story11,song12b}.  A cluster is included in the disturbed sample if \BCGd\  > 3$\Delta\theta$.  We find \NdisturbedSZonly\  systems that satisfy both conditions.

%

\subsubsection{X--ray cluster morphology: \Aphot}
The bulk of the merging cluster samples studied in this work are identified through the \BCGd\ selection.  We extend these samples using X--ray information. For the XVP sample, \cite{nurgaliev17} computed the photon asymmetry index \cite[\Aphot;][]{nurgaliev13}.  \Aphot\ measures the deviation from axisymmetry of the  X-ray emission. 

Using the \Aphot $>0.6$ for disturbed clusters, we find that   \NAphotDISTlimited\ systems that fall into this  category\footnote{An eighth cluster with a high A$_{\rm phot}$ value, SPT-CLJ2332-5053, is excluded from this selection.  Inspection of the {\it Chandra} image shows that the high A$_{\rm phot}$ value is due to contamination from  SPT-CLJ2331-5051, a companion cluster possibly in a  pre-merger stage.}, one is present in our BCG-SZ selection.  Several of \Aphot\ selected disturbed clusters also present clear substructures in the red galaxy density maps. Visual inspection of the DES images confirm that the position of the BCGs in these 6 clusters is close to the SZ centroid, explaining the low \BCGd. This may indicate a different merging phase in comparison to the \BCGd\ selected systems, a projection effect, or a case of a minor merger that manages to disturb the gas but not the BCG. 

\subsubsection{BCG-X--ray centroid/peak offset}

Using  \BCGXd\ $>$ 0.4 \Rtwo, and assuming  negligible X--ray centroid/peak positional uncertainty,   on {\it Chandra} data, we find 1 new system, SPT-CLJ0346-5439, a cluster with a \BCGd\ of 0.37\Rtwo.  In the case of the   XMM sub-sample,  we also  add \NXMMDISTcentroidlimitedNEW\ cluster, SPT-CLJ0253-6046 at a \BCGd\ of 0.36\Rtwo, to the disturbed sample.

\subsection{Relaxed sample}
\label{sec:coolcore}
To highlight the changes in the galaxy populations due to the dynamical state, we create a relaxed sample.  Due to the large SZ centroid positional uncertainty, the BCG-SZ-centroid offset is not a good proxy to create such a sample.  We turn here exclusively to the clusters with X-ray information.  

We define a system as relaxed if any of the following three conditions is met: i) the cluster has \Aphot$ \leq 0.1$; ii) the cluster is classified as having a cool-core \citep[systems with $K_0<$ 30 keV ${\rm cm}^2$;][]{mcdonald13}; or iii) the BCG satisfies \BCGXd\ $<$ 42(71) kpc to the X--ray peak (centroid) \citep{mann12}. 

\subsubsection{X--ray morphology: \Aphot\ }
Applying the \cite{nurgaliev17} selection for relaxed systems (A$_{\rm phot} \leq 0.1$) we find \NAphotRELAXlimited\ clusters that satisfy this condition.

\subsubsection{Cool-core clusters}
\label{subsubsec:coolcore} 
We use  cool core SPT-SZ clusters reported in \cite{mcdonald13}, which are defined as systems with $K_0<$ 30 keV ${\rm cm}^2$.  This results in \MDcoolcoreLim\ clusters within our redshift-mass limits.   These \MDcoolcoreLim\ clusters include \MDcoolcoreLimInAPHOT\  already selected by the A$_{\rm phot} \leq 0.1$ criterion.

\subsubsection{BCG-X--ray centroid/peak offset}
\label{subsec:bcgx} 
The X--ray centroid/peak to BCG offset is a well known predictor of the dynamical state of a cluster.  We use the X--ray centroid  and peak  to BCG distance to complement the the morphology and coolcore samples.  To select relaxed systems we follow \cite{mann12}; we use \BCGXd\ $<$ 42 kpc if we use the X--ray  peak, and \BCGXd\ $<$ 71  kpc if we use the X--ray  centroid. We add \NXMMCCcentroidlimitedNEW\ clusters from the XMM data and \ChandraNEWpeakRelaxedNEW\  from {\it Chandra} data to this sample. 

We note that the conditions described in  \S~\ref{subsubsec:coolcore} and \S~\ref{subsec:bcgx} also include clusters classified as disturbed; SPT-CLJ0102-4603, SPT-CLJ0252-4828, SPT-CLJ0346-5439, and SPT-CLJ2218-4519. These systems are excluded from both the relaxed and disturbed samples.
SPT-CLJ0102-4603 is identified as a cool-core cluster with $K_0$ = $10.3^{+6.1}_{-5.8}$ keV ${\rm cm}^2$, a \BCGXd $<$ 42 kpc to peak, but an \Aphot\  of $0.68^{+0.18}_{-0.17}$.  Visual inspection of the X--ray image of SPT-CLJ0102-4603 reveals some signs of asymmetry.  It is conceivable that the cluster suffered a recent minor merger, without disturbing the core. SPT-CLJ0252-4824 is selected as a relaxed system due to   \BCGXd~$=0.03$ kpc to the X--ray peak flux, but it is also classified as disturbed with an \Aphot\ of $0.97^{+0.34}_{-0.28}$. Visual inspection of the X--ray image reveals clear signs of merging activity, with the gas having a filamentary structure, while the candidate BCG is clearly dominant and unique. In the case of SPT-CLJ0346-5439, we find a cluster with a $K_0<$ 30 keV ${\rm cm}^2$, \BCGd~$=0.43$\Rtwo, and a large BCG-X--ray offset. X--ray images reveal an ellipsoidal photon distribution, not much different than other clusters, while having two dominant BCGs, a central one and our candidate 0.34 magnitudes brighter ($\sim 50$ times the photometric error). In the case of SPT-CLJ2218-4519, \Aphot\  defines it as a relaxed cluster ($0.08^{+0.05}_{-0.08}$) but has a large \BCGXd\ value ($0.54$\Rtwo). The X--ray imaging shows a regular photon distribution confirming the  \Aphot\ value.  Furthermore, a visual and catalog inspection of the candidate BCG and bright galaxies close to the X--ray emission reveals that there may be a potential error in the catalogue magnitudes and that the BCG may be miss-identified.

\begin{figure}
	\includegraphics[scale=0.66]{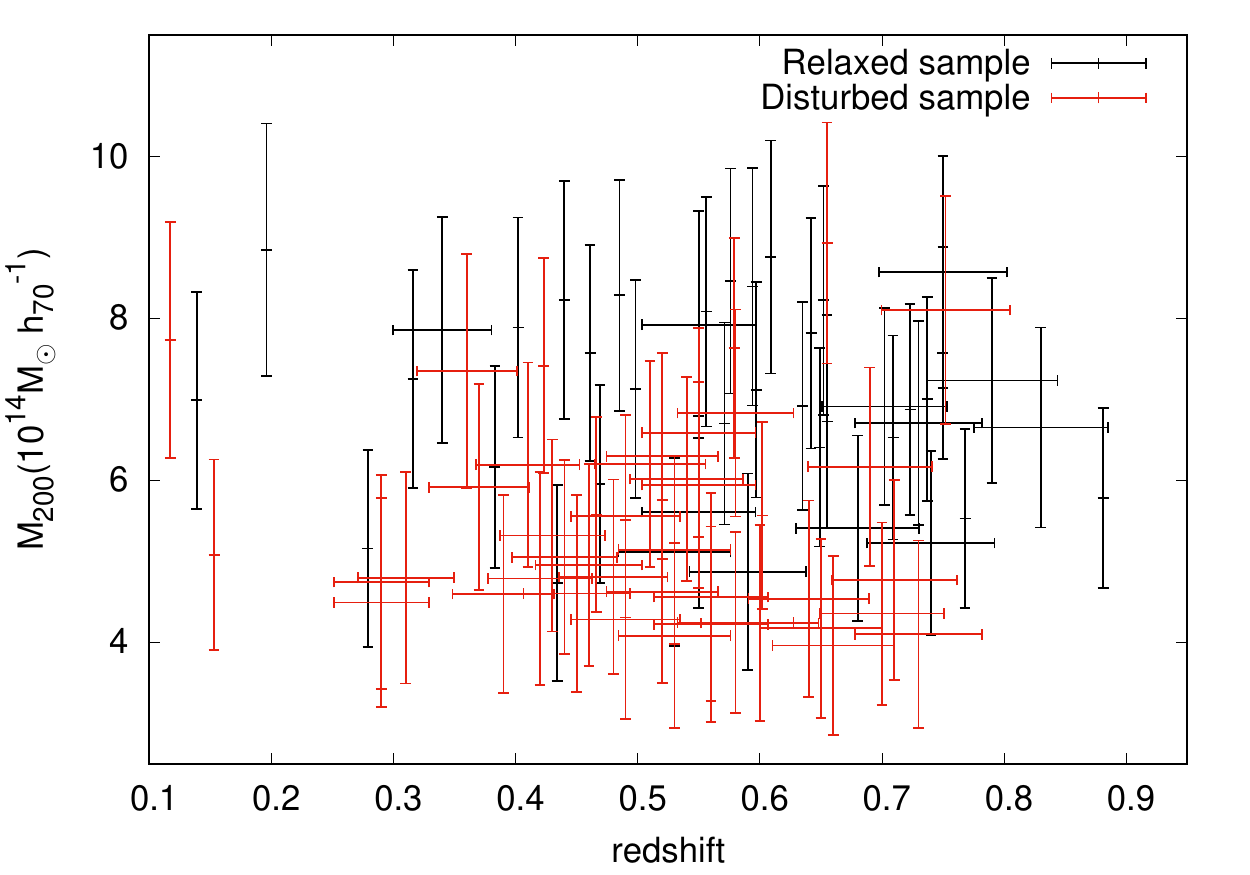}
    \caption{The location of relaxed (\Ncoolcore) and disturbed (\Ndisturbed) cluster sample mass-redshift plane. The mass and redshift cuts described in section \S~\ref{sub:zmass} 
    have been applied.  It can be seen that the average mass of the relaxed clusters are higher than the average mass of the disturbed clusters.} 
    \label{fig:massredshift}
\end{figure}

\begin{figure*}
	\centering
    \includegraphics[scale=0.67]{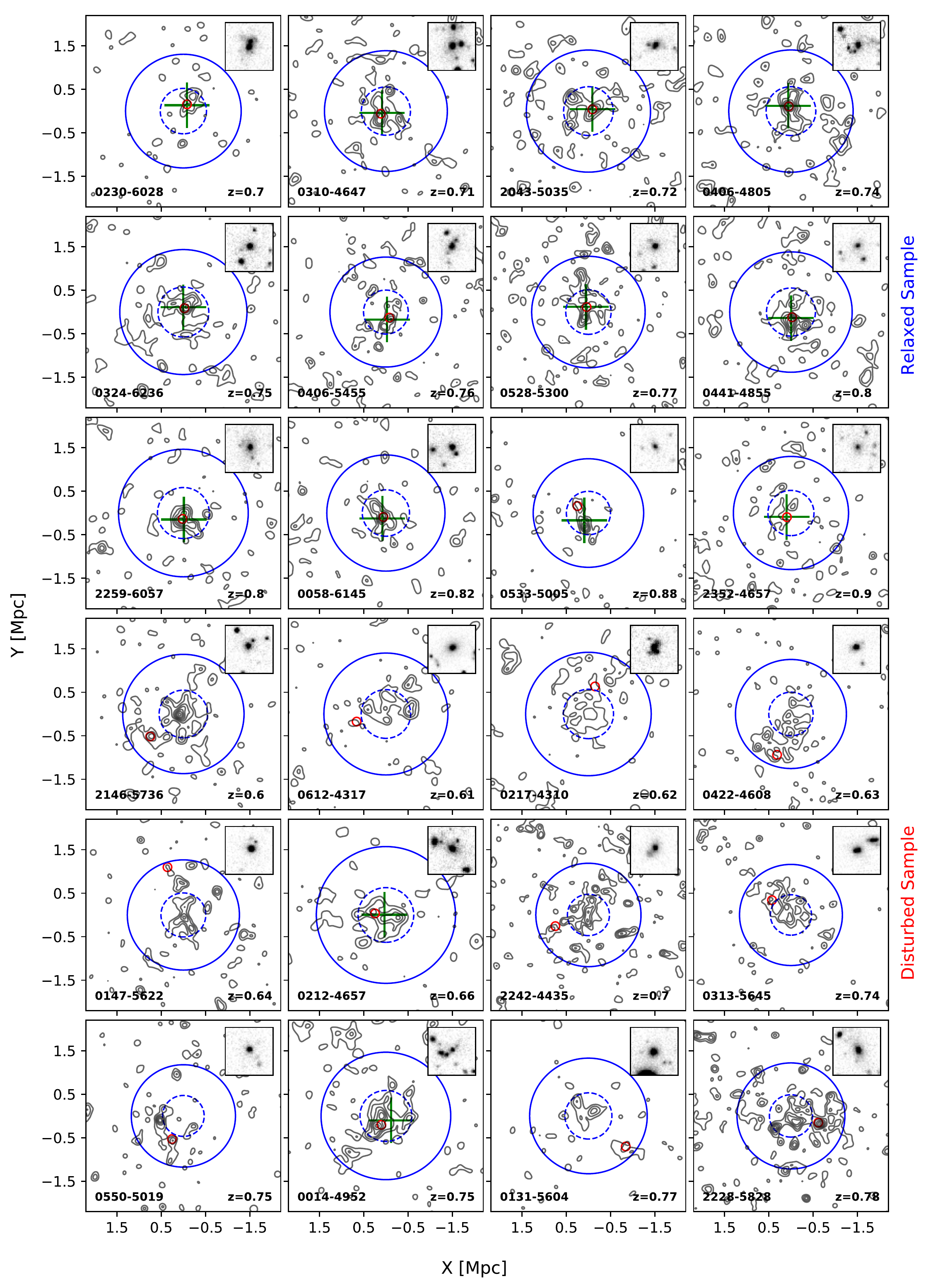}
    \caption{The top 12 clusters correspond to the highest redshift relaxed systems. The outer circled represents \Rtwo, while the inner circle corresponds to 0.4\Rtwo. Contours correspond to the red sequence galaxy distribution.  The green cross shows the X--ray peak emission while $i-$band images of the BCG is shown in the top right corner.  The relaxed systems are selected if the BCG is within 42/71 kpc of the X--ray peak/centroid \citep{mann12}, if $K_0<$ 30 keV ${\rm cm}^2$ \citep{mcdonald13}, or if it is classified as relaxed by its morphology with a A$_{\rm phot}<=0.1$ \citep{nurgaliev17}. The bottom 12 clusters correspond to the highest redshift disturbed systems centered in the SZ emission.  Most of the clusters have been selected by the large displacement of the BCG with respect to SZ center of X--ray peak/center, while a fraction relies on a gas morphology index  \Aphot.  In the latter case the BCG is not necessarily significantly displaced from the gas center, as shown by SPT-CLJ0014-4952.
    }
    \label{fig:optdisturbedrelaxed}
\end{figure*}

The final disturbed cluster sample is composed of  \Ndisturbed\ clusters (listed in Table~\ref{tab:disturbed}), while the final relaxed cluster sample has \Ncoolcore\   (listed in  Table~\ref{tab:relaxed}), with half of them at $z \gtrsim 0.6$, a redshift scarcely probed by previous studies. The mass and redshift distribution of both cluster samples, within the mass and redshift cuts described in \S~\ref{sub:zmass}, is shown in Fig.~\ref{fig:massredshift}.  Images for a sub sample of 12 disturbed systems and 12  relaxed systems  can be see in Fig.~\ref{fig:optdisturbedrelaxed}. 

We also create an intermediate sample of clusters, numbering  \NclustersMZcutNoUnrelaxedNorelaxed, consisting of all clusters that remain.  Note that such sample will contain disturbed and relaxed clusters (i.e., our disturbed and relaxed samples are by no means complete).

We also tested an optical proxy for the cluster dynamical state.  We used the \cite{wen15} recipe, but we were unable to find consistent results,  when comparing to samples created using X--ray data for dynamical state estimation.  We will explore other dynamical state proxies in  follow-up  papers.

\section{Galaxy populations}
\label{sec:galpop}

\subsection{Luminosity Function}
\label{sec:lf}

%
Following Z16, we construct individual cluster Luminosity Functions (LFs) using galaxies  within  a projected \Rtwo,  centered on the  BCG, and performing a statistical background subtraction  using the background region within 1.5 and 3\Rtwo.  
For  two populations; red sequence galaxies and all cluster galaxies, we use the same BCG to  set the bright limit of the LF, independently of its color. The projected, background-corrected LF is then  de-projected using an NFW profile with  a concentration of 3 ($c_{\rm  corr}$), a value in between the findings of Z16 and H17 for SPT clusters in different mass ranges. We use 0.5 magnitude bins for the LF, 
and scaled it by the cluster volume in Mpc$^3$. The BCG is excluded from the LF.

As in Z16 and H17, for all stars with a 2MASS J band magnitude $m_J < 14$, in both the cluster and background regions, we mask the area within a distance from the star depending on the star's magnitude\footnote{We use a radius = $10^{\rm (A_J+B_J \times J-band)}$ arcseconds, with $\rm A_J = 2.4378$, and $\rm B_J = -0.10576$.}.

Corrections due to masked regions and background over-subtraction (from cluster members at r $>$ 1.5 \Rtwo) are applied here as well.  In the  case of cluster  masked regions (within \Rtwo) we correct for the missing cluster galaxies using the NFW profile with the  concentration $c_{\rm  corr}$.  Also,  using the  same model,  we correct for the over subtraction due to cluster galaxies contaminating the  background  dominated  region (also corrected by masked areas).  This over subtraction can be expressed  mathematically in terms of the observables that we measure $\bar{N}^{1.5 < r'<3}$ and $\bar{N}^{r'<1}$, corresponding to the galaxies at  $1.5 < r' < 3$, and the galaxies within $r' < 1$, respectively,  where $r' = r / r_{200}$. Both cluster galaxies and background galaxies contribute to these numbers:
\begin{equation}
\label{equation1}
\bar{N}^{1.5 < r'<3} = N_{\rm back}^{1.5 < r'<3} + N_{\rm clus}^{1.5 < r'<3}
\end{equation}
and
\begin{equation}
\label{equation2}
\bar{N}^{r'<1} = N_{\rm back}^{r'<1} + N_{\rm clus}^{r'<1}
\end{equation}
Our first step is to create the LF by subtracting those two observables:
\begin{equation}
\label{eq:sub}
N_{\rm clus,LF}^{r'<1}= \bar{N}^{r'<1}  - A_{\rm N}\times  \bar{N}^{1.5 < r'<3} 
\end{equation}
Where $A_{\rm N}$ is the area normalization.  Replacing above equation with \ref{equation1} and \ref{equation2} we obtain:
\begin{displaymath}
N_{\rm clus,LF}^{r'<1}=  N_{\rm back}^{r'<1} + N_{\rm clus}^{r'<1}  - A_{\rm N}\times  (N_{\rm back}^{1.5 < r'<3} + N_{\rm clus}^{1.5 < r'<3}) 
\end{displaymath}
which can be simplified by canceling the background contribution ($N_{\rm back}^{r'<1} = A_{\rm N}\times  N_{\rm back}^{1.5 < r'<3}$): 
\begin{equation}
N_{\rm clus,LF}^{r'<1}= N_{\rm clus}^{r'<1}  - A_{\rm N}\times  N_{\rm clus}^{1.5 < r'<3} 
\label{equation3}
\end{equation}
This equation shows the over-subtraction term introduced in eq.~\ref{eq:sub}.  To correct for this we use a NFW profile to connect cluster galaxies at different radii:
\begin{displaymath}
N_{\rm clus}^{1.5 < r'<3} = \tau(c_{\rm corr}) \times N_{\rm clus}^{r'<1} 
\end{displaymath}
Using this relation in equation \ref{equation3} we obtain a coefficient we can use to correct such over-subtraction:
\begin{displaymath}
N_{\rm clus,LF}^{r'<1}=  N_{\rm clus}^{r'<1}   - A_{\rm N}\times   \tau(c_{\rm corr}) \times  N_{\rm clus}^{r'<1} 
\end{displaymath}
\begin{displaymath}
N_{\rm clus,LF}^{r'<1}=  N_{\rm clus}^{r'<1}  (1 - A_{\rm N}\times   \tau(c_{\rm corr}))
\end{displaymath}
\begin{displaymath}
 N_{\rm clus}^{r'<1}   = \frac{N_{\rm clus,LF}^{r'<1}}{(1 - A_{\rm N}\times   \tau(c_{\rm corr}))} = C \times N_{\rm clus,LF}^{r'<1}.
\end{displaymath}

The average correction $C$ is $1.06$ with a maximum of $1.085$.  The correction boosts the number of galaxies within $r_{200}$ to compensate for the  over-subtraction of cluster galaxies at $1.5 < r'<3$. We notice that these corrections are expected to be well defined for the relaxed systems as the galaxy distribution is expected to have a NFW shape.  In the case of the disturbed cluster sample, the suitability of the corrections are less clear.  If the clusters have a lower concentration due to a recent merger, such a correction would underestimate the contamination of cluster galaxies in the background area, underestimating the cluster richness.

\subsubsection{Completeness}
\label{sec:completeness}
The DES data used in this paper provides a homogeneous data set to a depth of \DESdepthgTen, \DESdepthrTen, \DESdepthiTen, and \DESdepthzTen~in $griz$ at 10 $\sigma$, as measured in our 84 clusters. For clusters at $z>0.45(0.65)$, the average photometry of \mstar+3(2) galaxies has a signal-to-noise ratio less than 10.  For these cases, completeness becomes important and we adopt correction factors to  enable analysis to a common depth relative to the cluster galaxy LF characteristic magnitude.  

%
The correction follows our previous work in \citet{zenteno11}:
we compare the galaxy number counts in the $griz$ bands in an annulus of radii 1.5-3 \Rtwo\ around each cluster, to the deeper Canada-France-Hawaii-Telescope  Legacy  Survey \citep[CFHTLS,][private   communication]{brimioulle08}\footnote{Count histograms correspond to the  D-1  1  \sqd\  field,  at l=  $172.0^{\circ}$  and  b  = $-58.0^{\circ}$  with a  magnitude limit  beyond r=27  and  a seeing  better  than  1.0\arcsec} by dividing the number count histogram in the cluster field by that in the CFHTLS field. If the ratio is well behaved (i.e., is stable around 1 before the completeness correction is needed, and the 10 $\sigma$ error in the catalog is consistent with the 90 \% completeness from the divided histograms), we fit the resulting  curve with an  error  function which  is used  to account for the  missing objects as we approach the 50\% completeness depth (no data is used beyond 50\% completeness limit).  If the curve is not well behaved, we use the data up to 90\% completeness, defined by the catalog errors. As noted in \S~\ref{sec:desimaging}, given the average photometric depth, this completeness correction allows us to reach $m^*+3$ up to redshift $0.61\pm0.04$, or $m^*+2$ up to redshift $0.83\pm0.06$.
%

\subsubsection{Stacked Luminosity Function}
\label{sec:stackedlf}

To stack the data we bring each individual LF, extracted to up to \mstar+3, to the same magnitude frame.  We do this by subtracting from the cluster LF bins, in apparent magnitudes,  the SSP model $m^*$ magnitude given the cluster redshift (see \S~\ref{sec:bcgselection}). 
Once  the data are brought  to this common frame,  they are stacked using the inverse variance weighted average method used in Z16:
\begin{displaymath}
\label{eq:stack}
N_{j}=\frac{\sum_i  N_{ij}^{z=0} / \sigma_{ij}^2}{\sum_i 1/\sigma_{ij}^2}
\end{displaymath}
where  $N_{ij}^{z=0}$  is  the  number  of  galaxies  per  cubic Mpc  per magnitude bin  at redshift  zero, in  the $j$th  bin corresponding  to the $i$th cluster and $\sigma_{ij}$  is the associated statistical Poisson error.   We  obtain  $N_{ij}^{z=0}$   by  correcting  it  by  the evolutionary    factor    $E^2(z)$,   where    $E(z)=\sqrt{\Omega_{\rm m}(1+z)^3+\Omega_\Lambda}$ \footnote{This  scaling   is  appropriate  for self-similar  evolution, where  the characteristic  density within  the cluster  virial region  will scale  with the  critical density  of the universe.  }.

The errors of the stacked LF are computed as
\begin{displaymath}
\delta N_{j}=\frac{1}{(\sum_i 1/\sigma_{ij}^2)^{1/2}}
\end{displaymath}

Once the  stacked binned LF is  constructed, we fit for the three parameters of the Schechter Function (SF) $\Phi^*$, $m^*$, and $\alpha$ \citep{schechter76},
\begin{displaymath}
\phi(m)=0.4 \ln(10)\ \Phi^*10^{0.4(m^*-m)(\alpha+1)}\exp(-10^{0.4(m^*-m)}).
\end{displaymath}

\subsubsection{LF robustness tests}
\label{sec:robustness}
We use the offset between the BCG and the gas center, as well as the gas morphology, to classify the cluster dynamical state, and use the BCG as the cluster center to study galaxy populations.  If galaxy transformation takes place in  clusters, as they merge and relax, evidence of such transformation should be detected even if we vary the cluster center.  To corroborate this, we run several tests on our samples.
We take the relaxed and disturbed samples and randomize the  cluster center position within the 0.4-1 \Rtwo\ range from the gas proxy, with a flat probability 600 times. In particular, we stack the LF of clusters under different configurations including; (A) randomizing the cluster center position from the X--ray peak/centroid for the relaxed sample, and (B) randomizing the cluster center position from the SZ centroid (a loose gas center proxy to compare to the relaxed sample) for the disturbed sample.   We also include another two cases for comparison;  (C) we stack the disturbed sample LF using the SZ centroid as a center, and (D) we stack the LF of the intermediate population, using the BCG as the cluster center. Case D  is used to highlight differences between the intermediate population and the two extremes we are probing.   We then create 600 stacked LF by randomly selecting  N unique clusters from this sample.  As the goal is to compare these randomised samples to the relaxed and disturbed samples, N corresponds to the number of clusters of the largest sample, relaxed or disturbed.

\subsubsection{Halo Occupation Number}

To estimate the richness of the clusters and compare the number of galaxies as a function of the dynamical state, we estimate the  Halo  Occupation  Number (HON).  We estimate the HON by integrating the Schechter Function.
\begin{displaymath}
N= 1 +  N^s, \ {\rm with}\ N^s =  V \phi^*\int^{\infty}_{y_{\rm low}} y^{\alpha}
e^{-y}\ dy
\end{displaymath}
where 1 accounts for the  BCG, which is not  part of the
LF fit, $V$  is the cluster  virial volume, $y_{\rm  low}=L_{\rm low}/L_*$, and  $\alpha$  and  $\phi^*$  are  obtained by fitting the Schechter function in  (\S~\ref{sec:stackedlf}). We integrate the LF to  $m^*+3$\footnote{Notice that on average, our individual luminosity functions depth allows us to reach $m^*+3$ only in clusters to redshift $0.61\pm0.04$, beyond that, we are fitting shallower LFs.} to compare to the work of \cite{lin04a}.

\section{results}
\label{sec:results}
\subsection{Brightest Cluster Galaxies}

\subsubsection{Membership tests}
\label{sub:BCGtests}
BCG selection is a key component of this work.  In \S~\ref{sec:bcgselection} we used the spectroscopic data of \BCGallMATCH\ systems to show that we expect a miss-classification rate of about 2\%.  This may vary with cluster sample type. For example, as the disturbed sample is defined by how far the BCG is from the SZ centroid, the likelihood of choosing a non-cluster galaxy may be higher. Here we look into the sub-samples in more detail.

We compare our selected SPT BCGs to candidates found in  \cite{mcdonald16}.  The selection of BCGs in \cite{mcdonald16}  has been done independently from ours, but the methodology is similar.    \cite{mcdonald16} selected the brightest red sequence galaxy within \Rtwo, but then visually inspected the images to change the candidate if any of the following conditions were met: i) if more than one dominant galaxy is found significantly  closer to the X--ray peak (29\% of cases), or if there is a blue galaxy close to the X--ray peak (3\% of cases).  %
Under those conditions we expect our disturbed sample to only contain BCGs that were not selected by the above procedure, while for the relaxed cluster sample the selection should be very similar.  
From \NChandralimited\  clusters in common between the DESY3-SPT sample and  \cite{mcdonald16}, we find 28 clusters in the relaxed sample and 5 in the disturbed sample.  From the relaxed sample we find 5 discrepancies.  A visual inspection reveals that the different choice of BCGs in \cite{mcdonald16} is indeed based on those cases where there is another bright/dominant galaxy close to the X--ray center. In all those cases (SPT-CLJ0033-6326, SPT-CLJ0232-5257, SPT-CLJ0334-4659, SPT-CLJ0533-5005, and SPT-CLJ2148-6116) our BCG selection is brighter, within 71\arcsec of the X--ray peak,  while the  \cite{mcdonald16} selection was even closer. From the disturbed sample we find that all the 5 selected BCGs are different from our selection while for the remaining 9 cluster matches with the intermediate cluster sample, a match between 5 BCGs is found.

We also repeat the BCG background estimation done for all \NclustersMZcut\ clusters in the DESY3-SPT sample (see \S~\ref{sec:bcgselection}) for the disturbed sample. We find that for 23\% of the clusters there are no RS galaxies brighter than the BCG in the 1-3 \Rtwo\ area. Within the same area, we find that for 34\% of the sample, there is only 1 RS galaxy brighter than the BCG. For the 23/13/7 \% of the sample, we find 2/3/4 galaxies brighter than the BCG.  Combining this information we get a probability of miss-identification of 14\%, the same when all \NclustersMZcut\ clusters are used.  We compare our disturbed cluster sample BCGs to the RM selection. A match between clusters renders 36 matches and 7 SPT clusters with no RM counterpart.  All the non-matches correspond to RM masked areas where a no cluster search was carried out.  When matching the BCG, we find that out of the 36 matches, only 5 correspond to our BCG selection.  For 16 cases, the RM BCGs correspond to a bright galaxy close to the SZ-centroid, while for 15 cases a different galaxy, much fainter, is chosen.

Finally, a cross match between the spectroscopic sample and the BCGs from the disturbed cluster sample renders 6 matches; 3 selected from X--ray information (SPT-CLJ0014-4952, SPT-CLJ0212-4657, and SPT-CLJ0551-5709), and 3 by \BCGd\ (SPT-CLJ0145-5301, SPT-CLJ0307-6225 and SPT-CLJ0403-5719).  All of them at the same redshift as their host cluster.  This provides confidence in our BCG selection.

\subsubsection{Brightness and color}
\label{sub:bcglum}
\begin{figure}
    \centering
    \includegraphics[scale=0.57]{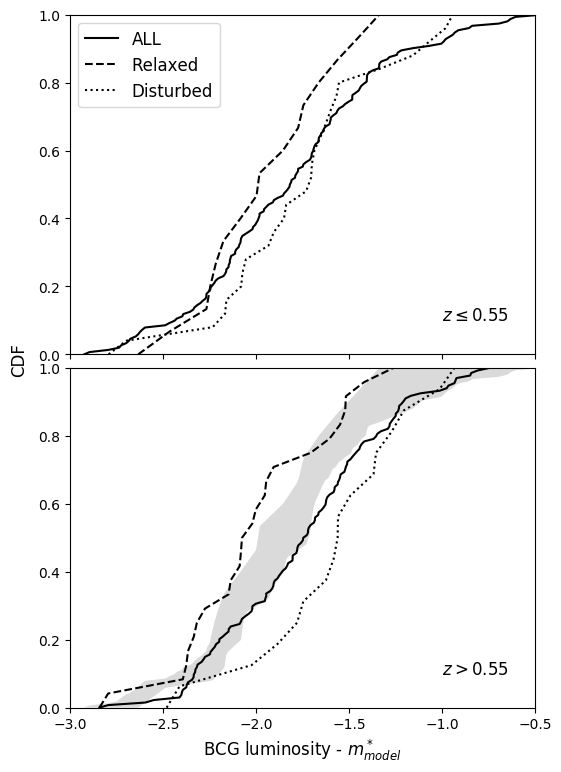}
    \caption{Cumulative distribution of the BCG luminosity ($r-$band at $z<0.35$, $i-$band at $z<0.7$, and $z-$band at $z>0.7$) for the total (solid line), relaxed (dashed line), and disturbed (dotted line) cluster samples. {\it Top panel}: BCGs at $z \leq 0.55$. {\it Bottom panel}: BCGs at $z > 0.55$. Shaded area corresponds to the area covered by the BCG distributions at the low redshift bin.}

    \label{fig:bcglum_highz}
\end{figure}

The BCGs have an average luminosity  of $m^*-1.7$, with 95\% of the population fainter than $m^*-2.45$, and all of them fainter than $m^*-2.8$.  We chose to limit the brightness of the BCG to $m^*-4$, and the observed distribution supports this choice.  The cumulative BCG luminosity distributions, with respect to the host cluster \mstar, for the disturbed, relaxed, and for all clusters are shown in Fig.~\ref{fig:bcglum_highz}.  The plots show that the BCG luminosity distribution of the samples agree at lower redshift (top panel), while they are somewhat different at higher redshift (bottom panel).  In the high redshift bin the BCGs in relaxed clusters are brighter than in disturbed clusters, and for all clusters combined.  At z $\leq 0.55$, a  Kolmogorov-Smirnov (KS) test shows that a comparison between the total and the relaxed sample renders a p-value of \KSRelaxedIntermediateLowz, while the p-value between the disturbed and relaxed sample is \KSMergingRelaxedLowz, confirming their agreement (or the lack of evidence of disagreement). At z $> 0.55$, the KS test renders a p-value of \KSRelaxedIntermediate\  between the total and the relaxed sample, and a p-value of \KSMergingRelaxed\ between the disturbed and relaxed sample. Both results reject the hypothesis that the distributions come form the same parent distribution. In general, there are several potential reasons for the difference in relaxed and disturbed systems. In relaxed clusters, cooling from the ICM will contribute to the BCG mass growth; if cool cores last for $\sim 10$ Gyr, BCGs would gain $\gtrsim10^{10} M_\odot$ from cooling even if it is highly suppressed by the AGN \citep[see][ and references therein]{mcdonald18}.  In regards to our samples, a factor to consider is that the BCG stellar mass correlates with halo mass \citep{brough08,lidman12}. As the average total mass of the relaxed sample is slightly higher than the average total mass of the disturbed clusters (see Fig~\ref{fig:massredshift}), one would expect BCGs in relaxed systems to be more luminous. In term of photometry,  deblending problems could also bias high the luminosity of BCGs in relaxed systems as they are located in crowded places. Nevertheless, the BCG luminosity of relaxed and disturbed samples in the $z < 0.55$ range agree, indicating that such effects must be of little importance. Understanding how those elements can play a different (larger) role at $z > 0.55$ is more  challenging.  We can speculate that at higher redshift galaxies sink more efficiently to the bottom of the potential, merging with the BCG and increasing its luminosity.  As the relevant dynamical friction timescales  depend on the distance between the BCG and other cluster galaxies, a factor to consider is the clusters size.  As clusters radius are a function of the critical density of the universe, they are more compact  at higher redshifts;   \Rtwo\ for a given cluster is smaller at high redshift than at low redshift  as \Rtwo $\propto 1.0/(\Omega_M(1+z)^3 + \Omega_\Lambda)^{2/3}$.  Having galaxies  closer together at higher redshift would make dynamical friction indeed more efficient. 

In the case of the disturbed sample,  BCGs may suffer the  loss of weakly bound stars during the merging event, although with a KS test p-value of \KSMergingIntermediate,  we can not reject the hypothesis that both the disturbed sample and the all clusters sample, come from the same parent distribution. 

In terms of color, we find no significant differences between the distributions above and below $z=0.55$.  For the \NclustersMZcut\ BCGs in the full DESY3-SPT sample, only \BlueTOTALBCGsPct\% of the BCGs are bluer than the red sequence (defined with a  width of $\pm$ 0.22 magnitudes; see \S~\ref{sec:bcgselection}). This fraction is lower than what has been reported in the literature. Based on SDSS DR6 data of over 14,300 BCGs in the 0.1-0.3 redshift range, \cite{pipino11}  reported  4 to 9\% of the BCGs being 0.3 magnitudes bluer than the $g-r$ red sequence. If we use 0.3 as a color limit, we only find that \BlueTOTALBCGsPIPINORANGEPct\% of all SPT BCGs studied here are blue.  This difference is alleviated if we factor in the cluster mass; indeed,  when a mass cut is applied ($\gtrsim 1.4 \times 10^{14} $M$_\odot$), \cite{pipino11} found a lower rate of 6\%.  Given that our systems are among the most massive clusters in the Universe, a \BlueTOTALBCGsPIPINORANGEPct\% rate of blue BCGs seems consistent with previous findings. 



Comparing the colors of the BCGs in the relaxed and disturbed samples we find no blue BCGs in the former.  The relaxed sample includes SPT-CLJ2043-5035, a cluster with excess blue emission in HST images and a star formation rate of $33.1^{+7.1}_{-4.6} M_\odot$yr$^{-1}$ \citep{mcdonald19}.  Our DES photometry shows a color of $0.17$ magnitudes bluer than the RS for  SPT-CLJ2043-5035, placing the BCG color within our 0.22 red sequence magnitude width, showing that the RS width also includes star forming BCGs. In the disturbed sample we find  \BlueDisturbedBCGs\ blue BCGs, about \BlueDisturbedBCGsPct\%, in the latter sample. It is worth noting that as the disturbed sample is defined by how far the BCG is from the SZ centroid, the likelihood of choosing a foreground (bluer) galaxy may be higher, although we showed in \S~\ref{sec:bcgselection} and  \S~\ref{sub:BCGtests} that overall  with our selection we expect a miss-classification rate of about 2 to 9\%. 

\begin{figure}
	\includegraphics[scale=0.55]{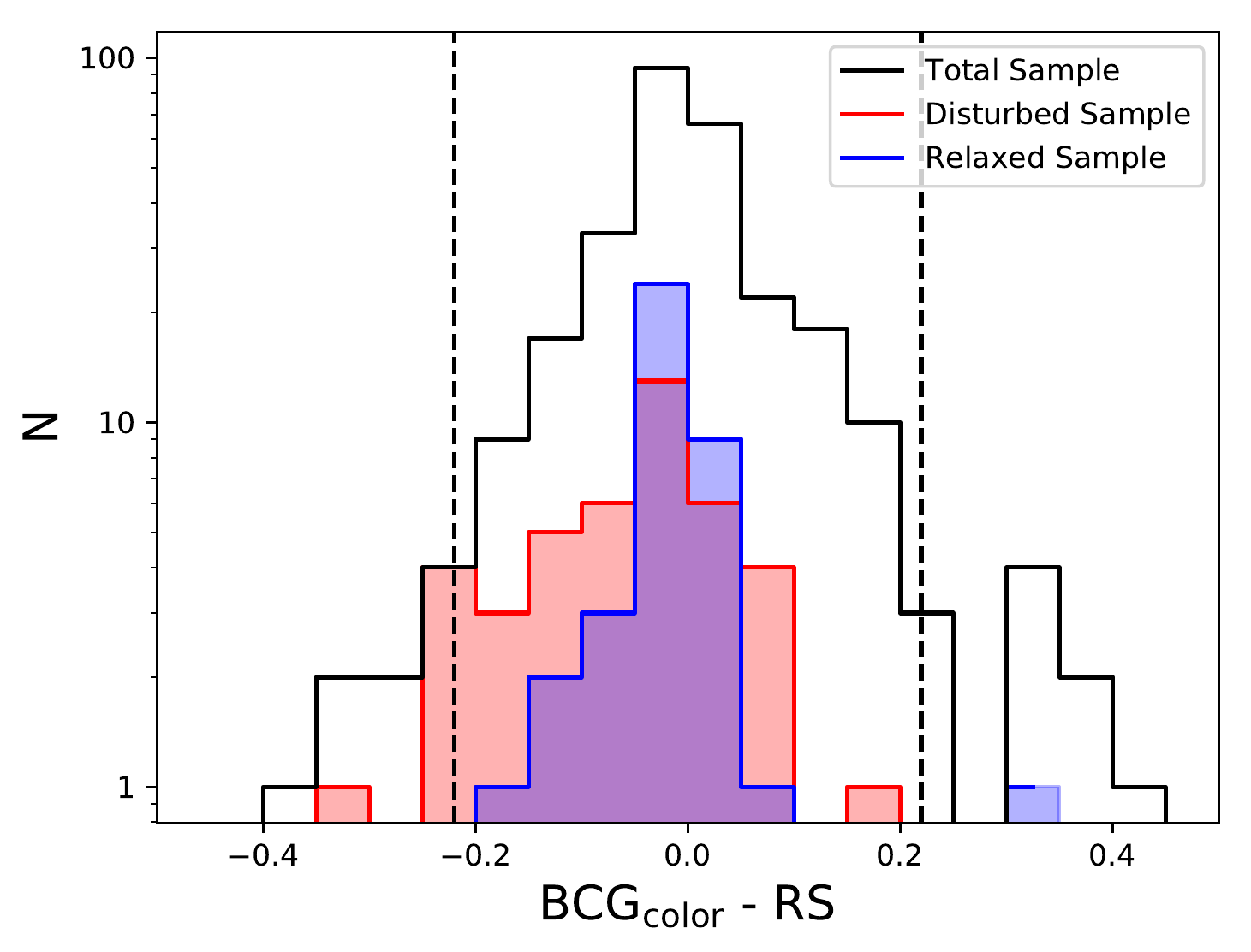}
    \caption{BCG color for the total, the relaxed, and the disturbed cluster samples with respect to the red sequence (as defined in \S~\ref{sec:bcgselection}). The BCG colors, and \mstar, are estimated depending on the cluster redshift; $g-r$ at $z<0.35$, $r-i$ at $z<0.7$, and $i-z$ at $z>0.7$. Vertical lines are at $\pm$ 0.22 mags, corresponding to our 3$\sigma$ red-sequence width \citep{lopez04}.}
    \label{fig:bcgcolor}
\end{figure}

In our relaxed sample the well known coolcore cluster SPT-CLJ2344-4243 \citep[aka the Phoenix cluster][]{mcdonald12a}, with a blue BCG with an extreme star formation rate, is missing. With a mass greater than $10^{15} $M$_\odot$, the Phoenix cluster is out of our mass-redshift limits.  The BCGs color distribution for the total, disturbed, and relaxed samples is shown in Fig.~\ref{fig:bcgcolor}.

\subsubsection{BCG-SZ offset distribution}
\begin{figure}
	\includegraphics[scale=0.41]{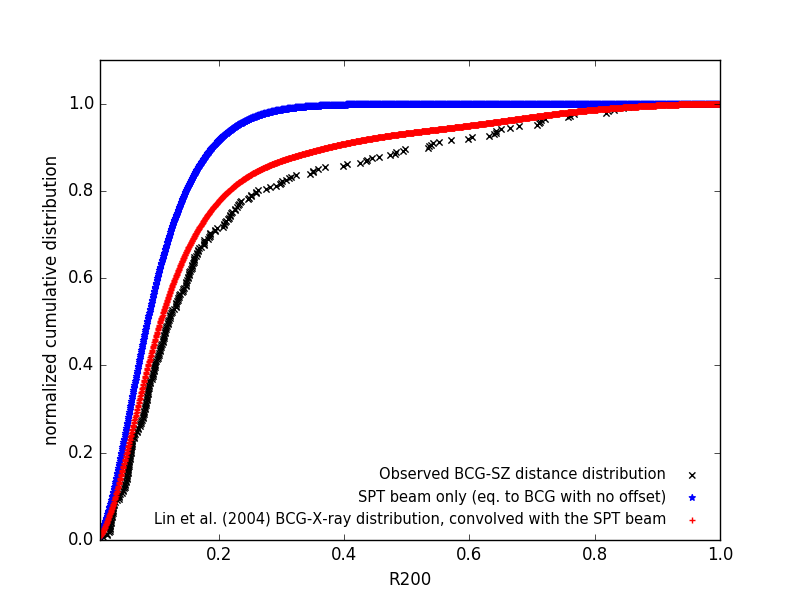}
    \caption{The full cluster sample cumulative central offset distribution  (black), \BCGd\, compared to an X-ray sample from the literature (red) and the case of pure SPT positional uncertainty (blue).}
    \label{fig:linconvolved}
\end{figure}

As the BCG-SZ offset distribution encodes information of the cluster sample's average dynamical state, we can compare it to other BCG clustercentric distance  distributions estimated for clusters  selected in a different way, and test the possibility that the SZ sample is, for example, more merger rich \citep{rossetti16,lovisari17,lopes18}.    In particular, we compare our SZ-selected cluster sample to an X--ray selected cluster sample.  
An X--ray study which performs the BCG selection in a similar fashion as us (see \S~\ref{sec:bcgselection}) is \citet[LM04 hereafter]{lin04b}. The LM04 sample consists of 93 galaxy clusters and groups at $z \le 0.09$, drawn from several studies for which $T_X$ has been measured.  The cumulative offset distribution for the DESY3-SPT sample, normalized to \Rtwo, and a comparison to the LM04 sample (convolved with the SPT-SZ positional uncertainty) are plotted in  Fig.~\ref{fig:linconvolved}.

The blue line shows the offset distribution that would be observed if all BCGs had actually no offset from the cluster center, because of the positional uncertainty of the SZ centroid due to the SPT beam. The black crosses show the \BCGd\ distribution measured in this work. The \BCGd\ distribution for the SPT sample studied here is consistent with LM04 results;  a Kolmogorov-Smirnov (K-S) test shows a p=\SPTLinPvalue\ (at D=\SPTLinDvalue\ \Rtwo) between the LM04 sample and our \BCGd\ distribution  providing no evidence that SPT-SZ selected clusters are richer in disturbed clusters than the LM04 sample, a result that  confirms our findings in \cite{song12b}.

The ideal comparison is between samples that cover the same redshift range with the BCGs selected in the same fashion.  In the LM04 case, the X--ray sample is selected in a rather inhomogeneous way and consists of low redshift systems.  On the  other hand the LM04 BCG selection is very similar to ours.  X--ray studies that span larger redshift ranges and have been used to compare to SZ samples are those of HIFLUGCS \citep{reiprich02}, REXCESS \citep{bohringer07}, and MACS \citep{ebeling01}.  In particular, \cite{rossetti16} showed discrepancies between the offset distribution of the Planck sample \citep{planck13-20} and the aforementioned X--ray surveys.  The difficulty with using such samples is that the BCG has been selected in an `inhomogeneous' fashion.  While \cite{rossetti16} highlights the differences between the Planck sample and the X--ray studies, it is also true that a comparison between some of those  very same X--ray samples, e.g., HIFLUGCS versus REXCESS, will also reject the hypothesis that those two offset distributions are drawn from the same parent distribution. This is an expected outcome from samples that have a different BCG selection function.



\begin{figure}
 	\includegraphics[scale=0.46]{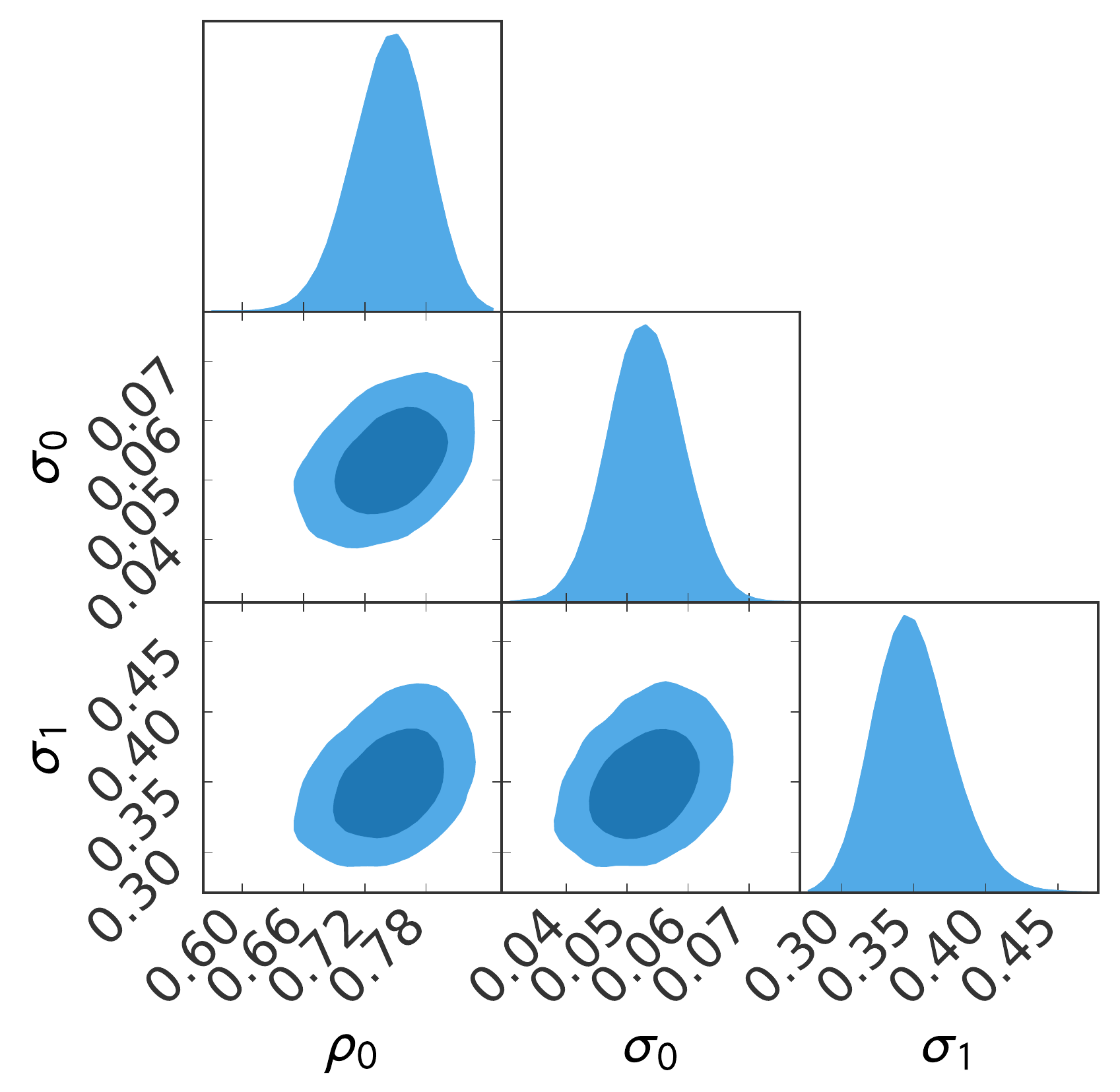}
    \caption{1$\sigma$ and 2$\sigma$ solutions of the three parameter model of equation~\ref{eq:gaussians} that  describes the \BCGd\  distribution.
    \label{fig:gaussiansposterior}
    }
\end{figure}

We  model the offset distribution using a sum of two distributions. One to describe a cluster population with small \BCGd,   and the second one to model the disturbed cluster population. Following \cite{saro15}, we use the equation:
\begin{equation}
\label{eq:gaussians}
P(x) = 2\pi x \left(\frac{\rho_0}{2\pi\sigma_0^2}e^{-\frac{x^2}{2\sigma_0^2}} + \frac{1-\rho_0}{2\pi\sigma_1^2}e^{-\frac{x^2}{2\sigma_1^2}}\right)
\end{equation}
where $x = r/$\Rtwo\ or $x = r/$\Rfive, $\rho_0$ is the fraction of the population more centrally located and with a variance of $\sigma^2_0$, while $1- \rho_0$ corresponds to the more perturbed  population, which has a variance of $\sigma^2_1$. 
We find that the observed  \BCGd\ can be described as a dominant population of clusters with smaller \BCGd ($\sigma_0$ (\Rtwo/\Rfive)  = \SigmaZeroRtwo /\SigmaZeroRfive), and a subdominant population of clusters with large \BCGd ($\sigma_1$ (\Rtwo/\Rfive) = \SigmaOneRtwo/\SigmaOneRfive).  The fit for the two distributions, as well as the sum, is shown in Fig.~\ref{fig:arcsecdistribution}. %
The marginalised posterior distributions for the three parameter model  can be seem in  Fig.~\ref{fig:gaussiansposterior}.  In comparison to the results obtained by \cite{saro15}, using redMaPPer selected central galaxies, we find that the parameters of the  population with small \BCGd\  agree, within the error bars. For the disturbed population we find a larger $\sigma_1$ value than the \cite{saro15} result ($0.25^{+0.07}_{-0.06}$). This is expected as the redMaPPer central galaxy selection is based on the galaxy luminosity, galaxy photometric redshift, and local galaxy density \citep{rykoff14}, while in our case we are after the brightest cluster member.  Furthermore, \citet{hoshino15} found that  20\% to 30\% of the RedMaPPer central galaxies are not the brightest cluster member. 
For this comparison we estimate the parameters in eq.~\ref{eq:gaussians}  using the \BCGd\ distribution as a function of \Rfive. 

\subsection{Luminosity Function}
\begin{figure}
	\includegraphics[scale=0.65]{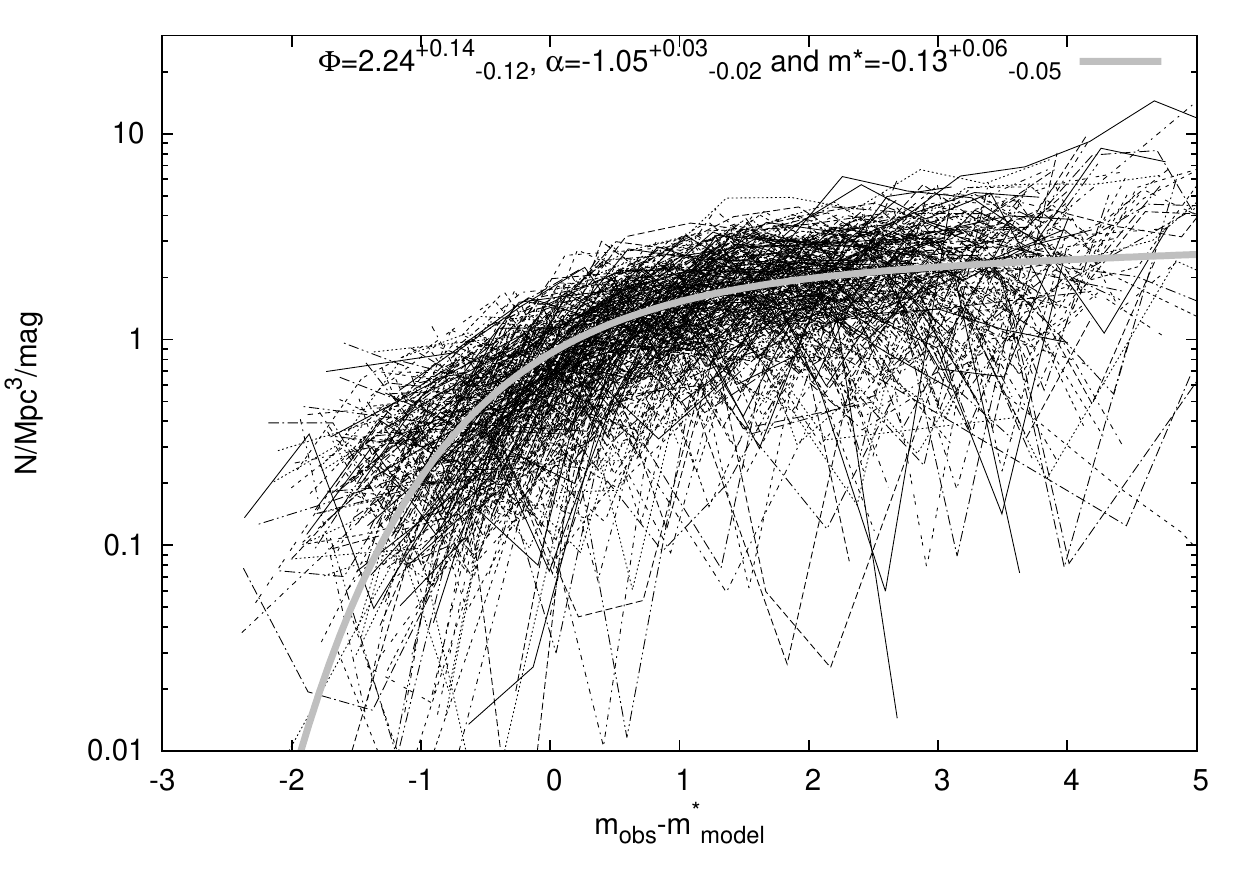}
\caption{Luminosity functions of individual clusters for the \allpop\ population up to their completeness limit. We use the \mstar\ model and the band redwards the 4000\AA\  break to offset all LFs to the same photometric rest-frame. The stacked result (gray solid line) is created using individual LFs up to \mstar\ + 3, or the individual completeness limit if \mstar\ + 3 can not be reached. }
    \label{fig:allstacked}
\end{figure}

We fit the Schechter function to the stacked LFs exploring several  configurations such as center (BCG or SZ), radius (1 or 0.5 \Rtwo), and population (all cluster galaxies included or just the red-sequence galaxies).   An example of individual LFs as well as a fit to the stacked result,   for the all cluster galaxies population within \Rtwo, is shown in Fig.~\ref{fig:allstacked} while the stacked LFs are shown for the relaxed and disturbed samples in Fig.~\ref{fig:stackedLFall}.
The results are summarized in table~\ref{tab:results} and shown in Fig.~\ref{fig:amresults_lowz}  and Fig.~\ref{fig:amresults_highz}.  We discuss them in the following sub-sections. 

\subsubsection{Faint end evolution}
\label{sub:alpha}

\begin{figure}
	\includegraphics[scale=0.65]
{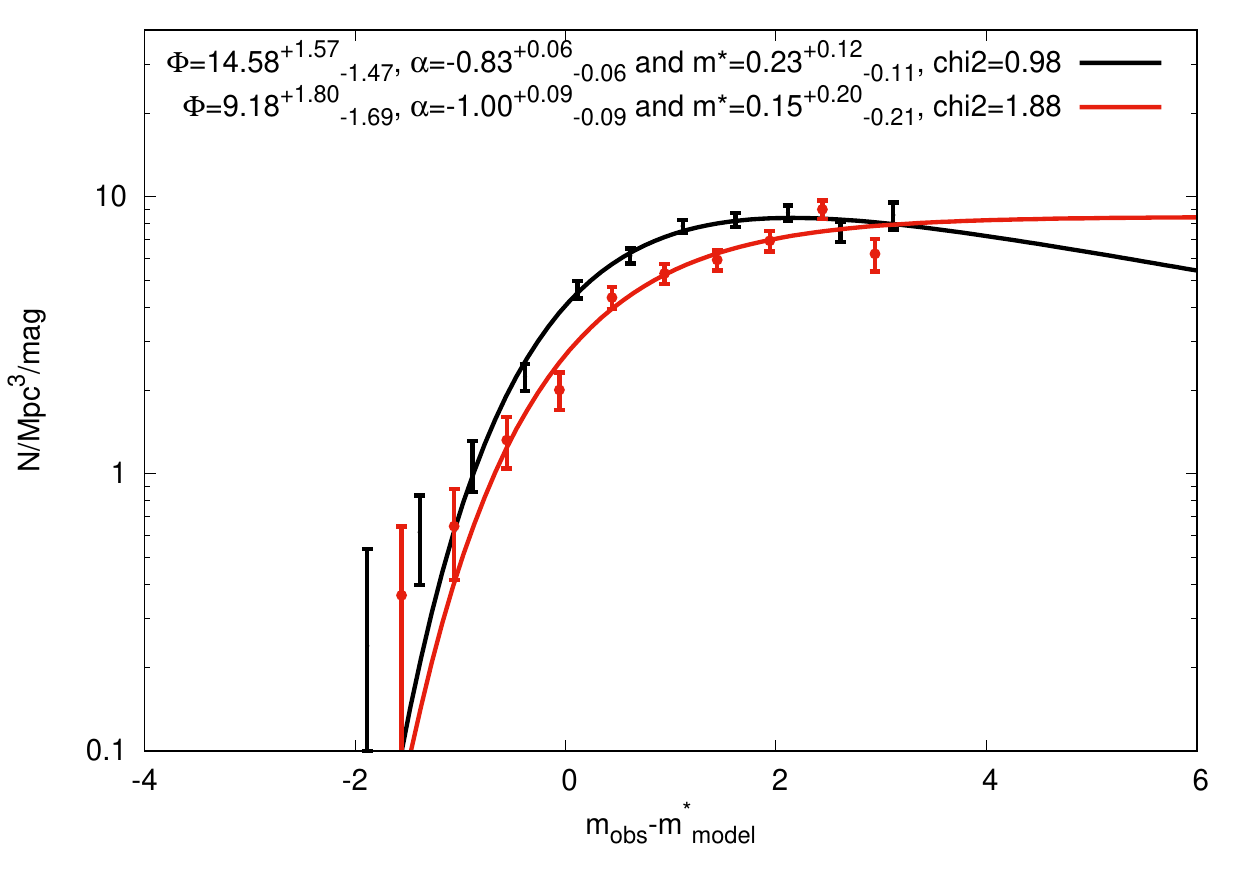}
    \caption{Stacked luminosity functions for the disturbed (red) and relaxed (black) sample, within $0.5 \times$ \Rtwo, including all cluster galaxies. We use the \mstar\ model and the band redwards the 4000\AA\  break to offset all LFs to the same photometric frame}
    \label{fig:stackedLFall}
\end{figure}

\begin{figure}
	\centering
	\includegraphics[scale=0.55]{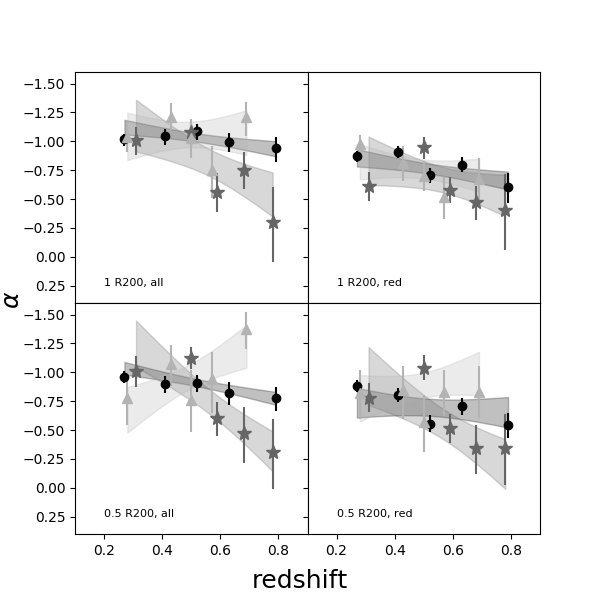}
    \caption{Faint end $\alpha$ evolution for the disturbed sample (triangles), the relaxed sample (stars), and all clusters excluding the disturbed and relaxed sample (dots). The shaded areas correspond to 1$\sigma$ contours.}
    \label{fig:alphaevolution}
\end{figure}

In general, we find differences between the faint end slope of the relaxed and disturbed samples, as exemplified in Fig.~\ref{fig:stackedLFall}. In Fig.~\ref{fig:alphaevolution} we explore the redshift evolution in the slope showing a distinct difference emerging at z $\gtrsim 0.6$, with a steeper $\alpha$ for the disturbed sample than for the relaxed sample. The redshift evolution figure indicates  that  the disturbed sample faint end slope is consistent with no evolution, while $\alpha$ for the relaxed clusters seems to evolve, becoming shallow at higher redshifts.  In the case of relaxed systems, the evidence for evolution is the strongest in the core of the clusters when all cluster galaxies are included ($2.7 \sigma$), and the weakest  at \Rtwo\ for red sequence galaxies ($1.0\sigma$). All evolution results are shown in Table~\ref{tab:alphaevolution}.

In the literature, only a few studies have sufficiently deep data to properly investigate the dependence (and evolution) of $\alpha$ on the cluster dynamical state, and all of them are at $z<0.6$.  \cite{barrena12} used a sample consisting of 5 radio-selected disturbed  and  5 cool-core clusters at $0.2 \lesssim z \lesssim 0.3$.  They find a steeper $\alpha$ for disturbed samples when the all cluster galaxies  within \Rtwo\ are included, but with large error bars which make them statistically indistinguishable from relaxed clusters.  \cite{depropris13} studied a sample of 11 disturbed clusters and 5 relaxed systems at $0.2 < z < 0.6$, classified according to X-ray, strong, and weak lensing information.  \cite{depropris13} again do not find differences between relaxed and disturbed clusters. These literature results, albeit using small samples, seem consistent with what we see at $z < 0.6$ in Fig.~\ref{fig:alphaevolution}, where $\alpha$ agrees between samples.

\begin{figure*}
     \includegraphics[scale=0.95]{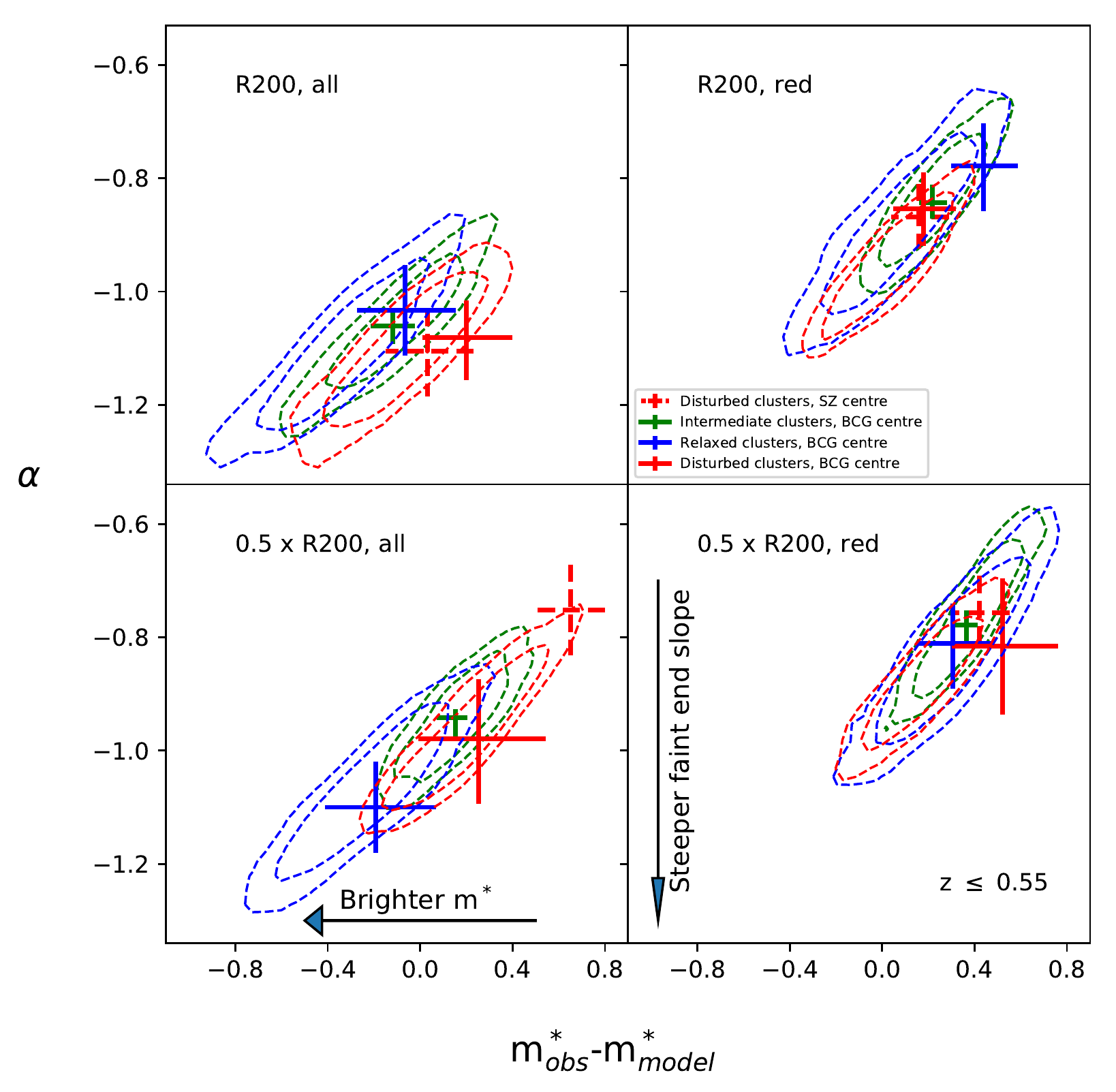}	
    \caption{$\alpha-\Delta m^*$(m$^*_{obs}-$m$^*_{model}$) for cluster galaxies (all) and red sequence galaxies (red), within 1 and 0.5 \Rtwo, at $z \leq 0.55$.  We obtain m$^*_{obs}$ by fitting the data, while  m$^*_{model}$ corresponds to the model as discussed in \S~\ref{sec:bcgselection}. Solid red crosses represent  disturbed clusters  using the BCG as a center (\NdisturbedLowZ\ clusters), dash red crosses represent  disturbed clusters  using the SZ centroid as a center, blue crosses correspond to the relaxed sample (\NcoolcoreLowZ\ clusters) while  green crosses correspond to the intermediate sample (see \S~\ref{sec:robustness} for details; \NAllexclLowZ\ clusters), both using the BCG as a center.   %
    Dashed red/blue contours correspond to 68\% and 90\% of the fits,  when the center is randomized 600 times within 0.4-1 \Rtwo\  of the SZ center, for the disturbed/relaxed sample.    Green contours correspond to 68\% and 90\%  of the fits, obtained by stacking \NdisturbedLowZ\ randomly selected clusters from the intermediate cluster sample, using the BCG as a center 600 times.  At this redshift range all cluster populations seems to occupy a very similar $\alpha$-$\Delta m^*$ parameter space.}
    \label{fig:amresults_lowz}
\end{figure*}

\begin{figure*}
    \includegraphics[scale=0.95]{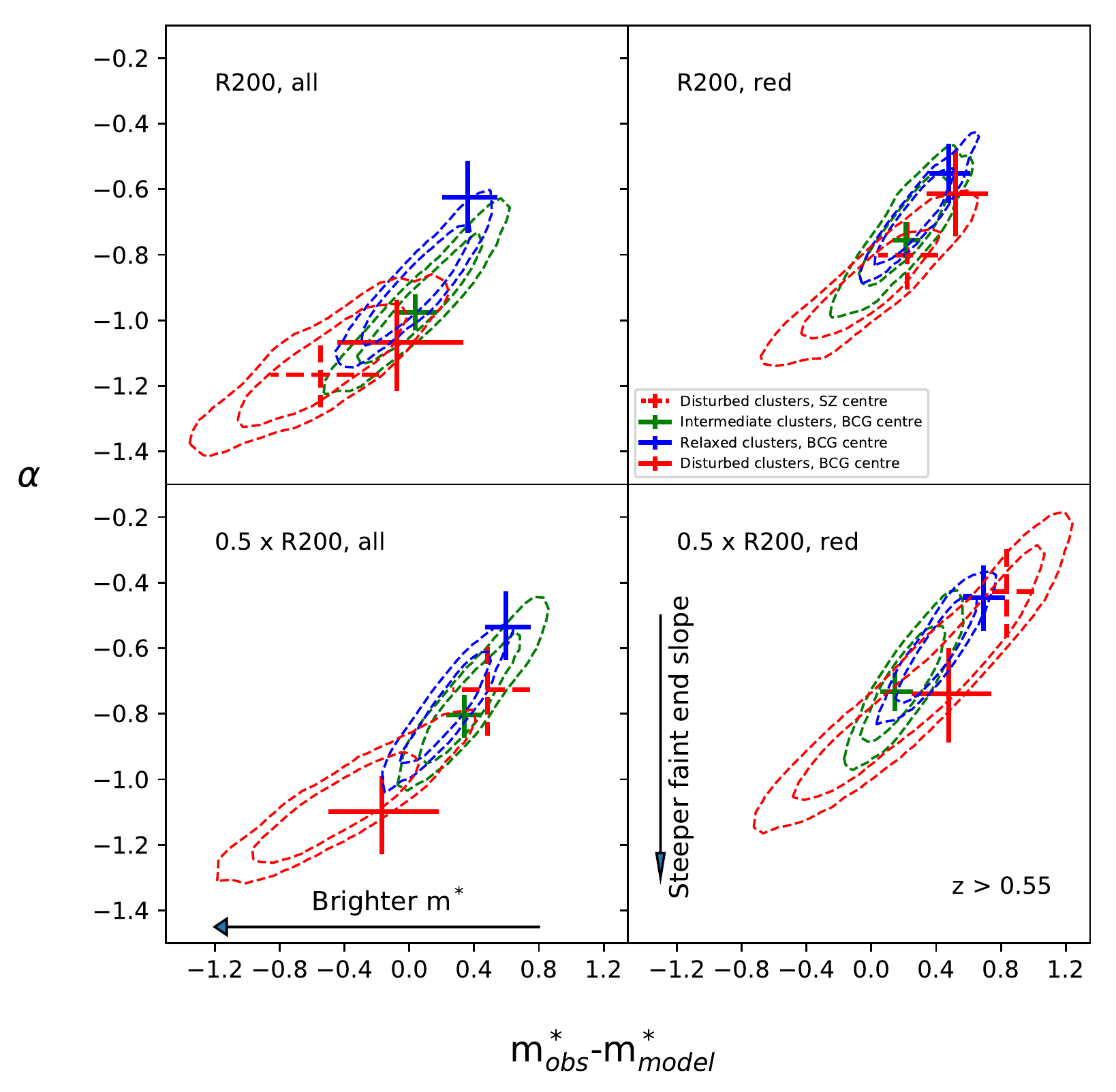}
    \caption{Similar to Fig.~\ref{fig:amresults_lowz} but for $z > 0.55$. The number of disturbed clusters in this redshift bin is \NdisturbedHighZ, while we find \NcoolcoreHighZ\ relaxed systems and \NAllexclHighZ\ clusters for the intermediate sample. The green curves correspond to stacking of \NcoolcoreHighZ\ systems instead of \NdisturbedLowZ, as in the $z\leq0.55$ bin. In this  redshift bin there is a tendency for the population in disturbed clusters to have a steeper $\alpha$ and a brighter \mstar.}

    \label{fig:amresults_highz}
\end{figure*}

To study in detail the effects of the cluster dynamical state on galaxies, we  split at $z=0.55$ the samples into two sets.  The redshift cut is motivated by the fact that at  around that redshift the LF parameters between the two samples start to differ,  with the lower redshift bin similar to the redshift range explored by literature work, and because  $z=0.55$ is the median redshift for the B15 parent sample.  
The low redshift sample results (\mstar\ and $\alpha$ Schechter parameters) are shown in Fig.~\ref{fig:amresults_lowz}, while the high  redshift sample results are shown in Fig.~\ref{fig:amresults_highz}.  For both figures the solid red crosses correspond to the BCG-centered stacked LF disturbed sample results, while the blue crosses correspond to the BCG-centered stacked LF relaxed sample results.   Fig.~\ref{fig:amresults_lowz} shows virtually no difference between the two samples, while Fig.~\ref{fig:amresults_highz} shows the disturbed sample having a steeper faint end slope than the relaxed sample for all the cases except for the red population within \Rtwo. 
For the higher redshift bin,  there is a tendency for disturbed and relaxed clusters to populate opposite extremes of the $m^*-\alpha$ plane, while the intermediate population (green crosses) populates the area in between the two aforementioned samples.  At $z>0.55$, this may point out to a continuous galaxy transformation as the  dynamical state  of the host cluster evolves, with the fainter population been accreted/destroyed as clusters relax.   A higher merging rate with the central BCG may be linked to the result, found in \S~\ref{sub:bcglum}, showing that BCGs in relaxed systems are brighter than BCGs in disturbed clusters as well as BCGs from the total \NclustersMZcut\  clusters sample.

The difference  between the lower and higher redshift samples might imply a larger galaxy merger rate, or galaxy destruction at higher redshift.  Studies measuring the galaxy merger rate in clusters seem to indicate a higher rate at high redshift than in the field. \cite{vandokkum99} used HST imaging of the massive cluster MS~1054$-$03 at $z = 0.83$  to show a higher merger rate for galaxies in this cluster in comparison to the field.  In terms of the dynamical state of MS~1054$-$03, \cite{tran99} conclude from the agreement between multiple mass proxies (dynamical, X--ray, and weak lensing) that it is a relaxed cluster, in spite of the irregular morphology of its galaxy distribution. Similar results have been found at even higher redshifts \citep[e.g.,][]{lotz13,watson19} but such studies have been limited to the bright end of the LF. To explain the change of the shape of the LF, the efficiency of merging or destroying  sub-$L^*$ galaxies must be higher at higher redshift.  Some hint of this can be found in studies dedicated to exploring the population of galaxies that contribute to the assembly of BCGs through cosmic time. \cite{burke13} explored the inner 50~kpc radius at redshift 0.8-1.3. They counted all galaxies up to 3.25 magnitude fainter than the BCG, which corresponds in average to a depth of $m^*+1.55$ for our BCG luminosity distribution, and found a larger number of large BCG companions than small galaxies companions.  
Similar results were found by \cite{lidman13} on a sample of 19 clusters at $0.84<z<1.46$.  \cite{lidman13} found an excess of bright companions (2nd to 4th brightest cluster members) within the 70 kpc inner annuli.
At $0.15 <z< 0.40$, exploring a distance between 30 to 50 kpc from the BCG, \cite{edwards12} found the opposite, a larger number of 1:10 and 1:20 BCG companions.  Although this may be what we are seeing, the radius we can resolve is about 10 to 20 times larger, and from dynamical friction arguments, galaxies at distances larger than $\sim$200 kpc, may take longer than the Hubble time to merge with the BCG \citep{burke13}.  Interestingly, \cite{birzan17}  arrived at a similar conclusion (an increase of  galaxy minor mergers at high redshift) by studying the radio luminosity evolution of BCGs (or central radio sources) in SPT clusters. To power the highest luminosity sources at $z>0.6$, a higher merging rate of gas rich galaxies with the BCG may be required.

A low redshift study that investigated the LF slope as a function of the cluster dynamical state is that of \citet[R13 hereafter]{ribeiro13}. R13 studied a sample of 183 clusters at $z \leq 0.1$, classifying them as relaxed (84\%) or disturbed (16\%) according to the Gaussian or non-Gaussian shape of the line-of-sight galaxy velocity distribution. They found a steeper LF faint end for relaxed clusters. As mentioned previously, if we use the lower redshift sample, closer to the range that R13 probes, we find $\alpha$ is consistent between the disturbed and relaxed samples (Fig.~\ref{fig:alphaevolution} and ~\ref{fig:amresults_lowz}). 
To test if the dynamical state proxy may play a role, we use \Nspectocompare\ SPT clusters with spectroscopy from \cite{capasso19} and \cite{bayliss16}, and classify their galaxy velocity distribution as Gaussian (\NspecG\ clusters at $0.30<z<0.77$) or non-Gaussian (\NspecNG\ clusters at $0.48<z<0.75$) using the Anderson-Darling test. Stacking the LFs of the spectroscopically classified samples shows a steeper $\alpha$ for disturbed clusters, although with larger error bars.  Results are shown in Table~\ref{tab:results}.

\begin{table}
\scriptsize
	\centering
	\caption{Result from a linear fit to the data separate in 5 redshift bins.  The center used is the BCG.}
	\label{tab:alphaevolution}
	\begin{tabular}{lccccc} 
    \hline \hline 
		Sample & intercept & slope & significance & R/R$_{200}$ & Gal. type\tablefootnote{``all" corresponds to all cluster galaxies, while ``red" corresponds to red sequence galaxies} \\
		\hline
All & $-1.24^{+0.11}_{-0.12}$ & $0.46^{+0.24}_{-0.24}$ & 1.92 & 1 & all \\ 
Intermediate & $-1.22^{+0.12}_{-0.13}$ & $0.36^{+0.26}_{-0.27}$ & 1.33 & 1 & all \\ 
Unrelaxed & $-0.82^{+0.35}_{-0.35}$ & $-0.37^{+0.70}_{-0.70}$ & 0.53 & 1 & all \\ 
Relaxed & $-1.53^{+0.34}_{-0.33}$ & $1.27^{+0.64}_{-0.65}$ & 1.95 & 1 & all \\ 
\hline
All & $-1.04^{+0.12}_{-0.12}$ & $0.54^{+0.25}_{-0.25}$ & 2.16 & 1 & red \\ 
Intermediate & $-0.96^{+0.14}_{-0.15}$ & $0.38^{+0.31}_{-0.31}$ & 1.23 & 1 & red \\ 
Unrelaxed & $-0.78^{+0.37}_{-0.38}$ & $-0.05^{+0.77}_{-0.76}$ & 0.06 & 1 & red \\ 
Relaxed & $-1.03^{+0.32}_{-0.33}$ & $0.64^{+0.62}_{-0.63}$ & 1.02 & 1 & red \\ 
\hline
All & $-1.12^{+0.13}_{-0.12}$ & $0.48^{+0.26}_{-0.26}$ & 1.85 & $\sfrac{1}{2}$ & all \\ 
Intermediate & $-1.17^{+0.14}_{-0.13}$ & $0.50^{+0.30}_{-0.29}$ & 1.72 & $\sfrac{1}{2}$ & all \\ 
Unrelaxed & $-0.46^{+0.48}_{-0.48}$ & $-1.19^{+0.91}_{-0.90}$ & 1.31 & $\sfrac{1}{2}$ & all \\ 
Relaxed & $-1.86^{+0.36}_{-0.37}$ & $1.98^{+0.73}_{-0.73}$ & 2.71 & $\sfrac{1}{2}$ & all \\ 
\hline
All & $-0.95^{+0.14}_{-0.14}$ & $0.50^{+0.29}_{-0.30}$ & 1.67 & $\sfrac{1}{2}$ & red \\ 
Intermediate & $-0.78^{+0.16}_{-0.16}$ & $0.16^{+0.34}_{-0.35}$ & 0.46 & $\sfrac{1}{2}$ & red \\ 
Unrelaxed & $-0.70^{+0.54}_{-0.53}$ & $-0.46^{+1.05}_{-1.05}$ & 0.44 & $\sfrac{1}{2}$ & red \\ 
Relaxed & $-1.47^{+0.38}_{-0.39}$ & $1.62^{+0.78}_{-0.78}$ & 2.08 & $\sfrac{1}{2}$ & red \\ 
\hline
		\hline
	\end{tabular}
\end{table}
%

\subsubsection{\mstar\ evolution}

Using the BCG as the cluster center, we find that \mstar\ is consistent across the samples for all cases in the lower redshift bin.  For the higher redshift bin, there is a tendency for the disturbed samples to have a brighter \mstar\ than for the relaxed case, which is more  significant when all cluster galaxies are used.  The fainter \mstar\ in relaxed clusters can be understood as the result of the merging of bright satellites with the BCGs as the cluster relaxes, changing the shape of the LF bright end.  Such a picture is consistent with brighter and more massive BCGs in relaxed clusters as discussed in ~\S~\ref{sub:bcglum}.

Literature works on $z<0.6$ samples, such as \cite{barrena12} and \cite{depropris13}, show that the \mstar\ of relaxed and disturbed sample are in agreement  across populations and radii, while works such as R13 ($z \leq 0.1$) and \cite{wen15} find contradictory results.  \cite{wen15} used an optically selected cluster sample of 2092 systems \citep{wen13}, at 0.05 $< z <$ 0.42,  and they used the galaxy distribution of galaxies to classify the dynamical state of the clusters. \cite{wen15} used $r_{500}$ and \Rtwo\ for the radius, the BCG for the cluster center, and 
galaxies with photometric redshifts within 0.04 of the cluster redshift.
%
They  found that \mstar\ is 0.27 mags. brighter for the disturbed sample, an effect that is most important within $r_{500}$, with no contribution in the area between $r_{500}$ and \Rtwo.  On the other hand, R13 found that relaxed clusters have a brighter \mstar\ than the disturbed sample.  
A difference between R13, \cite{wen15} and other studies (including ours) is the dynamical state proxy.  While the relaxed systems in \cite{barrena12} and \cite{depropris13} are classified using gas proxies, \cite{wen15} and R13 are classified using the shape of the galaxy distribution or the shape of the velocity dispersion, respectively. Different proxies may work best at different merger states and can be prone to different merger geometry.  Our method is prone to select mergers in the plane of the sky,  selection based on the deviation from a aussian shape of velocity dispersion are better at identifying mergers along the line-of-sight. In principle, a combination of effects could conspire to produce different results.  We will explore this aspect in future work. 

\subsubsection{Robustness of the LF results}

The robustness results are represented as dashed curves in Fig.~\ref{fig:amresults_lowz} and Fig.~\ref{fig:amresults_highz}.  While for the sample at $z<0.55$ all curves agree, differences arise in the higher redshift range. For the $z>0.55$ bin, Fig.~\ref{fig:amresults_highz}, the relaxed sample with a randomised center (case A; in blue) shows that, as expected, as we move away from the cluster core, more faint galaxies are included in the LF, i.e., we retrieve a steeper faint-end slope.  The 68\% and 90\%  intervals shown in Fig.~\ref{fig:amresults_highz} overlap partially with the LF of the disturbed sample that uses the BCG as the center (solid red crosses).   This result alone implies that we cannot rule out, at a statistically significant level, the possibility that the disturbed sample  can be described as the relaxed sample with a perturbed center, instead of a family of clusters with transformed galaxy population due to the merging process. To further explore differences between the relaxed and disturbed samples, we randomise the center of  the disturbed sample in the same fashion (case B; red contours in Fig~\ref{fig:amresults_highz}). The distribution of case A solutions tend to occupy a different region in the $\alpha$ and \mstar\ plane,  than for case B, supporting the case that differences in galaxies can be found in clusters with different dynamical states, most significantly when all galaxies are considered.

If the SZ center is used to stack the LF for disturbed clusters (case C; dashed red cross)  we observe a behaviour somewhere between the disturbed and relaxed populations. 

As a further comparison, we examine the intermediate population.  
%
Shown in green in Figs.~\ref{fig:amresults_lowz} and ~\ref{fig:amresults_highz} is case D, where we draw \NdisturbedLowZ/\NcoolcoreHighZ\ clusters without repetition  
from the low-z/high-z intermediate population, using the BCG as the cluster center, and creating stacked LFs.  Compared to the disturbed and relaxed samples, they tend to populate areas in between the relaxed and disturbed BCG-centered results.  This is again expected if disturbed and relaxed clusters are two extremes of the average population.

All these results combined illustrate a picture that supports a continuous galaxy transformation as clusters merge and relax. We highlight that this description is more significant when all galaxies are considered.  For red sequence galaxies, it seems that this population is rather impervious to the cluster dynamical state, at least within the redshift and mass range, and for the number of clusters here probed.  

\subsubsection{Halo Occupation Number}
The difference in the faint end slope of the LF found between the disturbed and relaxed cluster populations indicates that a transformation is at play as clusters evolve.     To measure the impact of this change on the number of galaxies, we estimate the HON for the different samples.  We have integrated the luminosity function to $m^*_{model}+3$ for the entire sample, and for the relaxed and disturbed samples to within 1 and 0.5 \Rtwo\ for all cluster galaxies  and red sequence galaxy only, within the whole redshift range. We use the individual cluster HON results, obtained by fixing \mstar\ to a model and fitting for $\alpha$ and $\phi$, the normalization of the HON versus mass relation, while keeping the slope fixed at $\gamma$=0.87 \citep{lin04a}.  We find a normalization N=\HONNORMLINPIVOTALLCLSthreeSigmaALL\ 
for all clusters,  lower but consistent with the value found by  \cite{lin04a} for X--ray clusters at 36$\pm$9.  
The disturbed sample has a normalization of \HONNORMLINPIVOTunrelaxedALLthreeSigma, while the relaxed one has \HONNORMLINPIVOTrelaxedALLthreeSigma.  
In the case of 0.5~\Rtwo, we find \HONNORMLINPIVOTALLCLSthreeSigmaHalf, \HONNORMLINPIVOTunrelaxedALLthreeSigmahalf, and \HONNORMLINPIVOTrelaxedALLthreeSigmahalf\  for the entire cluster sample, and for the disturbed and relaxed clusters, respectively.
For the red sequence galaxies we found normalizations for 1/0.5~\Rtwo\ of \HONNORMLINPIVOTALLCLSREDthreeSigmaALL/\HONNORMLINPIVOTREDCLSthreeSigmaHalf, \HONNORMLINPIVOTunrelaxedREDthreeSigmaALL/\HONNORMLINPIVOTunrelaxedREDthreeSigmahalf, and \HONNORMLINPIVOTrelaxedREDthreeSigmaALL/\HONNORMLINPIVOTrelaxedREDthreeSigmahalf\ for the entire cluster sample, and for the disturbed and relaxed clusters, respectively. All errors quoted are at the 3$\sigma$ level. 
Above results shows that, with the current samples, and photometric depth,  we are unable to constrain the HON with enough precision to find further evidence of the impact of the cluster dynamical state on the cluster's galaxies.

\section{Conclusions}
\label{sec:conclusions}
We have used imaging from the first three years of DES to find the BCGs, within \Rtwo\ of the centres of \NclustersMZcut\ SPT SZ-selected clusters.
%
%
We have  examined the offsets between the SPT-SZ centroid position and the BCG. Our findings show that we cannot reject the hypothesis that the SZ-selected  radial offset distribution is similar to that seen in an X--ray-selected cluster sample at low-z, provided the BCG is selected in the same fashion.  
Our findings confirm the results in \cite{song12b} providing no evidence that SPT SZ-selected cluster samples include a significantly higher fraction of  mergers than X-ray-selected cluster samples.  
%
We also showed that a key condition for this comparison to be meaningful is a sample of BCGs selected in the same manner.  To characterize the BCG-SZ offsets of the relaxed and disturbed clusters, we modeled the offset distribution as a sum of two distributions.  We found that the \BCGd\ distribution with small offsets agrees with the distribution shown in \cite{saro15}, based on redMaPPer selected central galaxies, while we find that the subdominant population of large \BCGd\ offsets is larger in our sample.  This is expected, as the only constraint on our BCG selection is for it to be the brightest within \Rtwo, while the selection of  redMaPPer central galaxies is more complex.  In fact, \cite{hoshino15} reported that 20\% to 30\% of the redMaPPer central galaxies are not the brightest cluster member, which is consistent with our results.

To select the most disturbed systems we use three proxies, finding \Ndisturbed\ clusters: the distance between the BCG and the SZ centroid $D_{\rm BCG-SZ}$, the distance between the BCG and the X--ray centroid/peak \BCGXd, and a X--ray morphology index \Aphot.  Within a similar redshift and mass range, we also use three proxies to select the most relaxed systems: XVP coolcore clusters, and clusters with low \BCGXd\ and \Aphot. We have found \Ncoolcore\ relaxed clusters. We compare the galaxy populations in these two extreme  dynamical states.  
We have built the stacked luminosity functions for all cluster galaxies and, separately, the red sequence galaxies, within  0.5 \Rtwo\ and \Rtwo, using the BCG as a proxy for the center.  We find that the LF of disturbed clusters have a steeper faint end slope $\alpha$ and brighter $m^*$ than the LF of relaxed clusters, although errors are large (see Table~\ref{tab:results}).  
%

We explored the redshift evolution of $\alpha$ and \mstar, dividing the different samples into 5 redshift bins  (Fig.~\ref{fig:alphaevolution}). We find that most of the difference between the disturbed  and relaxed clusters  comes from the clusters at $z \gtrsim 0.6$.  To explore the effects of the observed evolutionary trends in $\alpha$ and \mstar, as well as on the BCG, we separated the samples into two redshift bins at $z=0.55$. The low redshift bin (Fig.~\ref{fig:amresults_lowz}) shows no discernible difference between the disturbed, relaxed, and the intermediate clusters.  On clusters in the high redshift bin (Fig.~\ref{fig:amresults_highz}), clear differences can be seen between the disturbed and relaxed clusters when all galaxies are used.  The difference between the galaxy population in relaxed and disturbed samples is less clear for the red sequence population.  This is confirmed by tests we use to assess the robustness of our results to the cluster center.    
A possible explanation is that galaxies are accreted more efficiently at $z> 0.55$ by the BCG, as the host cluster dynamically relaxes.  This would explain the shallower $\alpha$, the fainter \mstar, and the brighter BCGs in relaxed clusters seen in Fig~\ref{fig:bcglum_highz}.  Fig~\ref{fig:bcglum_highz} shows the BCG cumulative brightness distribution for the relaxed, disturbed, and intermediate cluster samples for $z\leq 0.55$ and $z>0.55$, showing that although there is an agreement at lower redshift, BCGs in relaxed clusters in the high redshift bin are brighter.

Previous work on the study of the LF as a function of the cluster dynamical state has only been done at $z\lesssim 0.6$.  At that redshift range, our work shows agreement on the Schechter parameters between the relaxed and disturbed clusters.  When compared to literature work, the largest  discrepancy comes from a study on cluster samples built using the velocity distribution as the dynamical state proxy.  To understand such differences we explore the velocity dispersion method over the \Nspectocompare\ clusters in this sample with Gemini/GMOS and VLT/FORS2 spectroscopy.   We use the Anderson-Darling test to classify the cluster dynamical state, finding \NspecG\ relaxed and \NspecNG\ disturbed clusters.  We found that the disturbed stacked LF shows a steeper faint end than the relaxed stacked LF, albeit with large error bars.  We will explore this further with recently acquired Gemini/GMOS spectroscopic data for 14 disturbed clusters from this paper.

We also examined the Halo Occupation Number to explore the effect of the dynamical state of the clusters on the number of galaxies, per cluster mass.  We find an agreement within 3$\sigma$ between the total, relaxed and the disturbed clusters, but with large error bars for the later two samples.  If we investigate the HON redshift evolution, with the current sample size, we find no evidence of evolution for the relaxed and disturbed clusters. A larger sample is required to investigate this aspect.

A potential source of error is the survey depth.  As DES is a survey with fixed exposure times, high redshift clusters have the faint end of the luminosity function under-sampled in comparison to the low redshift sample.  This  feature translates to larger errors in  $\alpha$.  Also, the errors in color of the red-sequence selected galaxies become comparable to the RS (fixed) width; at redshift  $\sim 0.75/0.9$ the distribution of 68/50\% of the sub-$L^*$ galaxies, in the disturbed and relaxed clusters, have errors of $\sim 0.22$ or less.  The  increasing RS scatter as a function of magnitude potentially places faint galaxies out of the RS, producing an artificially  shallower $\alpha$.  While this may explain what we see in the relaxed sample, it does not explain the nearly constant $\alpha$ for the disturbed sample. Furthermore, if this effect would be a driving factor, we would expect a much stronger signal when red sequence galaxies are used than when all cluster galaxies are used, which is not seen. 
On the other hand, if the shallower $\alpha$ in relaxed systems is a real effect and not a product of faint photometry, then the missing sub-$L^*$ galaxy population must be being destroyed or going elsewhere, one of obvious candidates being the central galaxy.  As we mentioned above, the BCG luminosity at $z>0.55$ in relaxed systems are indeed brighter than BCGs in the disturbed and the intermediate clusters.  Furthermore, at $> 1$ magnitude brighter than the 10$\sigma$ depth across the cluster fields, BCG photometry is unaffected by the survey photometric limits.


In a follow up paper, we will expand this work using deeper $iz$ data from DES Y6, as well as dedicated pointed observations of $z > 0.7$ clusters. Furthermore, in the near future, eROSITA will  not only provide confirmation of the dynamical state of these clusters, but will render a sample of thousands of clusters that  will allow us to test to a higher significance the findings shown here, when combined with a BCG selection in a systematic way.

\begin{table*}
\begin{threeparttable}
	\centering
	\caption{Disturbed DESY3-SPT cluster sample.}
	\label{tab:disturbed}
	\begin{tabular}{lcccccccc} 
	    \hline
       \hline
        name & SPT R.A. & SPT Decl. & BCG R.A. & BCG  Decl. & redshift & M$_{200}$ & R$_{200}$ & offset \\
	    \hline
         & (J2000.0) & (J2000.0) & (J2000.0) & (J2000.0) & & $10^{14} h_{70}^{-1} M_\odot$ & [$^\prime$] & R$_{200}$ \\
	    \hline
SPT-CLJ0014-4952 &    3.6969 & $-49.8772$ &  3.70284 & $-49.88481^a$ & 0.752 & 8.10 & 3.27 & 0.10 \\
SPT-CLJ0038-5244 &    9.7204 & $-52.7390$ &  9.74027 & $-52.76584$ & 0.42 & 4.79 & 4.16 & 0.42 \\
SPT-CLJ0107-4855 &   16.8857 & $-48.9171$ & 16.86943 & $-48.88841$ & 0.60 & 4.24 & 3.08 & 0.60 \\
SPT-CLJ0111-5518 &   17.8446 & $-55.3138$ & 17.89407 & $-55.31981$ & 0.56 & 4.23 & 3.23 & 0.53 \\
SPT-CLJ0131-5604 &   22.9331 & $-56.0821$ & 22.87732 & $-56.10802$ & 0.69 & 6.17 & 3.17 & 0.77 \\
SPT-CLJ0135-5904 &   23.9753 & $-59.0814$ & 23.99831 & $-59.10447$ & 0.49 & 4.28 & 3.57 & 0.44 \\
SPT-CLJ0144-4807 &   26.1795 & $-48.1281$ & 26.10096 & $-48.13661$ & 0.31 & 4.80 & 5.27 & 0.60 \\
SPT-CLJ0145-5301 &   26.2645 & $-53.0295$ & 26.49096 & $-53.18341$ & 0.117 & 7.73 & 14.25 & 0.87 \\
SPT-CLJ0147-5622 &   26.9652 & $-56.3779$ & 26.99103 & $-56.33408$ & 0.64 & 4.54 & 3.01 & 0.92 \\
SPT-CLJ0151-5654 &   27.7898 & $-56.9110$ & 27.67812 & $-56.88732$ & 0.29 & 4.75 & 5.54 & 0.71 \\
SPT-CLJ0152-5303 &   28.2342 & $-53.0540$ & 28.26770 & $-53.08393$ & 0.55 & 6.59 & 3.79 & 0.57 \\
SPT-CLJ0212-4657 &   33.1061 & $-46.9502$ & 33.12022 & $-46.94882^a$ & 0.655 & 8.93 & 3.71 & 0.07 \\
SPT-CLJ0217-4310 &   34.4138 & $-43.1819$ & 34.40557 & $-43.15638$ & 0.52 & 6.30 & 3.89 & 0.40 \\
SPT-CLJ0253-6046 &   43.4619 & $-60.7725$ & 43.45072 & $-60.74986$ & 0.45 & 4.60 & 3.90 & 0.45$^b$ \\
SPT-CLJ0256-5617 &   44.0997 & $-56.2980$ & 44.08792 & $-56.30314^a$ & 0.58 & 6.83 & 3.70 & 0.07 \\
SPT-CLJ0257-4817 &   44.4463 & $-48.2970$ & 44.48667 & $-48.33896$ & 0.46 & 4.95 & 3.93 & 0.76 \\
SPT-CLJ0257-5842 &   44.3934 & $-58.7107$ & 44.42008 & $-58.66096$ & 0.44 & 5.05 & 4.09 & 0.76 \\
SPT-CLJ0304-4748 &   46.1503 & $-47.8115$ & 46.16431 & $-47.78205$ & 0.51 & 6.20 & 3.93 & 0.47 \\
SPT-CLJ0307-6225 &   46.8336 & $-62.4327$ & 46.84936 & $-62.40283^a$ & 0.579 & 7.63 & 3.84 & 0.48 \\
SPT-CLJ0313-5645 &   48.2620 & $-56.7548$ & 48.29124 & $-56.74206$ & 0.66 & 3.96 & 2.82 & 0.44 \\
SPT-CLJ0337-4928 &   54.4573 & $-49.4738$ & 54.43516 & $-49.51382$ & 0.53 & 5.14 & 3.59 & 0.71 \\
SPT-CLJ0337-6300 &   54.4692 & $-63.0103$ & 54.53705 & $-63.05146$ & 0.48 & 4.81 & 3.77 & 0.82 \\
SPT-CLJ0354-5904 &   58.5612 & $-59.0733$ & 58.61676 & $-59.09708$ & 0.41 & 6.19 & 4.61 & 0.48 \\
SPT-CLJ0403-5719 &   60.9681 & $-57.3237$ & 60.91901 & $-57.28362$ & 0.466 & 5.58 & 4.05 & 0.71 \\
SPT-CLJ0422-4608 &   65.7490 & $-46.1436$ & 65.76715 & $-46.18128$ & 0.70 & 4.36 & 2.79 & 0.85 \\
SPT-CLJ0429-5233 &   67.4315 & $-52.5609$ & 67.39019 & $-52.54447$ & 0.53 & 4.08 & 3.32 & 0.54 \\
SPT-CLJ0439-5330 &   69.9290 & $-53.5038$ & 69.90064 & $-53.53884$ & 0.43 & 5.32 & 4.23 & 0.55 \\
SPT-CLJ0451-4952 &   72.9661 & $-49.8796$ & 72.96825 & $-49.93968$ & 0.39 & 4.60 & 4.34 & 0.83 \\
SPT-CLJ0522-5026 &   80.5159 & $-50.4394$ & 80.54907 & $-50.41640$ & 0.52 & 4.62 & 3.51 & 0.53 \\
SPT-CLJ0526-5018 &   81.5087 & $-50.3147$ & 81.53626 & $-50.27385$ & 0.58 & 4.25 & 3.15 & 0.85 \\
SPT-CLJ0550-5019 &   87.5504 & $-50.3236$ & 87.56539 & $-50.34343$ & 0.65 & 4.17 & 2.90 & 0.46 \\
SPT-CLJ0551-5709 &   87.9041 & $-57.1557$ & 87.88328 & $-57.16026^a$ & 0.423 & 7.42 & 4.79 & 0.09 \\
SPT-CLJ0600-4353 &   90.0614 & $-43.8879$ & 90.07561 & $-43.93144$ & 0.36 & 7.35 & 5.40 & 0.50 \\
SPT-CLJ0611-4724 &   92.9212 & $-47.4111$ & 92.86097 & $-47.40201$ & 0.49 & 5.56 & 3.90 & 0.64 \\
SPT-CLJ0612-4317 &   93.0249 & $-43.2992$ & 93.06164 & $-43.30627$ & 0.54 & 6.02 & 3.73 & 0.44 \\
SPT-CLJ2040-5342 &  310.2194 & $-53.7116$ & 310.24179 & $-53.74783$ & 0.55 & 5.94 & 3.66 & 0.63 \\
SPT-CLJ2140-5331 &  325.0330 & $-53.5178$ & 325.01766 & $-53.48353$ & 0.56 & 4.56 & 3.31 & 0.64 \\
SPT-CLJ2146-5736 &  326.6957 & $-57.6148$ & 326.75290 & $-57.63592$ & 0.602 & 5.57 & 3.36 & 0.66 \\
SPT-CLJ2228-5828 &  337.2153 & $-58.4686$ & 337.17177 & $-58.47420$ & 0.71 & 4.77 & 2.85 & 0.49 \\
SPT-CLJ2242-4435 &  340.5195 & $-44.5897$ & 340.55949 & $-44.59951$ & 0.73 & 4.10 & 2.66 & 0.68 \\
SPT-CLJ2254-5805 &  343.5895 & $-58.0851$ & 343.39591 & $-58.10217$ & 0.153 & 5.08 & 9.75 & 0.64 \\
SPT-CLJ2344-4224 &  356.1481 & $-42.4100$ & 356.10884 & $-42.46793$ & 0.29 & 4.49 & 5.44 & 0.71 \\
SPT-CLJ2358-6129 &  359.7075 & $-61.4862$ & 359.70022 & $-61.43302$ & 0.37 & 5.92 & 4.92 & 0.65 \\
		\hline
	\end{tabular}
    \begin{tablenotes}
      \small
      \item {\bf Notes.} $^a$Clusters with a BCG candidate in \cite{mcdonald16}. Due to the \cite{mcdonald16} selection procedure, the BCGs presented for these 5 cases are different than what is reported here. See~\S~\ref{sec:bcgselection}  for details. $^b$The offset reported corresponds to the distance between the BCG and X--ray centroid.
    \end{tablenotes}
  \end{threeparttable}
\end{table*}

\begin{table*}
\begin{threeparttable}
	\centering
	\caption{Relaxed DESY3-SPT cluster sample.}
	\label{tab:relaxed}
	\begin{tabular}{lcccccccc} 
	    \hline
       \hline
        name & SPT R.A. & SPT Decl. & BCG R.A. & BCG  Decl. & redshift & M$_{200}$ & R$_{200}$ & offset \\
	    \hline
         & (J2000.0) & (J2000.0) & (J2000.0) & (J2000.0) & & $10^{14} h_{70}^{-1} M_\odot$ & [$^\prime$] & R$_{200}$ \\
	    \hline
SPT-CLJ0000-5748 &    0.2499 & $-57.8064$ &  0.25015 & $-57.80931^a$ & 0.702 & 6.91 & 3.25 & 0.01 \\
SPT-CLJ0033-6326 &    8.4767 & $-63.4463$ &  8.50134 & $-63.43077^b$ & 0.597 & 7.12 & 3.67 & 0.02 \\
SPT-CLJ0058-6145 &   14.5799 & $-61.7635$ & 14.58415 & $-61.76683^a$ & 0.83 & 6.65 & 2.87 & 0.03 \\
SPT-CLJ0123-4821 &   20.7923 & $-48.3588$ & 20.79565 & $-48.35626^a$ & 0.655 & 6.73 & 3.38 & 0.01 \\
SPT-CLJ0200-4852 &   30.1436 & $-48.8757$ & 30.14204 & $-48.87127^a$ & 0.498 & 7.13 & 4.18 & 0.01 \\
SPT-CLJ0230-6028 &   37.6418 & $-60.4689$ & 37.63540 & $-60.46297$ & 0.68 & 5.41 & 3.06 & 0.02 \\
SPT-CLJ0231-5403 &   37.7768 & $-54.0563$ & 37.77771 & $-54.06232$ & 0.59 & 4.87 & 3.26 & 0.03 \\
SPT-CLJ0232-5257 &   38.1876 & $-52.9578$ & 38.17795 & $-52.95628^b$ & 0.556 & 8.08 & 4.03 & 0.22 \\
SPT-CLJ0243-5930 &   40.8615 & $-59.5124$ & 40.86283 & $-59.51725^a$ & 0.635 & 6.92 & 3.48 & 0.00 \\
SPT-CLJ0257-5732 &   44.3506 & $-57.5426$ & 44.33722 & $-57.54837$ & 0.434 & 4.73 & 4.04 & 0.02 \\
SPT-CLJ0307-5042 &   46.9516 & $-50.7071$ & 46.96054 & $-50.70122^a$ & 0.55 & 7.92 & 4.03 & 0.04 \\
SPT-CLJ0310-4647 &   47.6291 & $-46.7834$ & 47.63545 & $-46.78566^a$ & 0.709 & 6.53 & 3.17 & 0.02 \\
SPT-CLJ0317-5935 &   49.3216 & $-59.5851$ & 49.31598 & $-59.59140$ & 0.469 & 5.96 & 4.12 & 0.01 \\
SPT-CLJ0324-6236 &   51.0530 & $-62.6021$ & 51.05100 & $-62.59883^a$ & 0.750 & 7.57 & 3.21 & 0.01 \\
SPT-CLJ0334-4659 &   53.5464 & $-46.9932$ & 53.52817 & $-46.98379^b$ & 0.485 & 8.29 & 4.48 & 0.24 \\
SPT-CLJ0342-5354 &   55.5220 & $-53.9118$ & 55.51920 & $-53.92059$ & 0.53 & 5.11 & 3.58 & 0.02 \\
SPT-CLJ0343-5518 &   55.7617 & $-55.3032$ & 55.75973 & $-55.31195$ & 0.55 & 5.61 & 3.59 & 0.04 \\
SPT-CLJ0352-5647 &   58.2367 & $-56.7996$ & 58.23958 & $-56.79772^a$ & 0.649 & 6.41 & 3.34 & 0.02 \\
SPT-CLJ0406-4805 &   61.7275 & $-48.0866$ & 61.73024 & $-48.08254^a$ & 0.737 & 7.01 & 3.16 & 0.01 \\
SPT-CLJ0406-5455 &   61.6906 & $-54.9210$ & 61.68571 & $-54.92573$ & 0.74 & 5.23 & 2.86 & 0.05 \\
SPT-CLJ0439-4600 &   69.8087 & $-46.0142$ & 69.80806 & $-46.01358$ & 0.34 & 7.86 & 5.77 & 0.03 \\
SPT-CLJ0441-4855 &   70.4511 & $-48.9190$ & 70.44958 & $-48.92335^a$ & 0.79 & 7.23 & 3.05 & 0.02 \\
SPT-CLJ0509-5342 &   77.3374 & $-53.7053$ & 77.33914 & $-53.70351^a$ & 0.461 & 7.57 & 4.52 & 0.01 \\
SPT-CLJ0528-5300 &   82.0196 & $-53.0024$ & 82.02214 & $-52.99814^a$ & 0.768 & 5.53 & 2.84 & 0.01 \\
SPT-CLJ0533-5005 &   83.4009 & $-50.0901$ & 83.41447 & $-50.08449^b$ & 0.881 & 5.78 & 2.64 & 0.29 \\
SPT-CLJ0542-4100 &   85.7167 & $-41.0044$ & 85.70855 & $-41.00012^a$ & 0.642 & 7.82 & 3.60 & 0.01 \\
SPT-CLJ0559-5249 &   89.9251 & $-52.8260$ & 89.93006 & $-52.82419^a$ & 0.609 & 8.76 & 3.88 & 0.04 \\
SPT-CLJ2011-5725 &  302.8527 & $-57.4217$ & 302.86241 & $-57.41971^a$ & 0.279 & 5.16 & 5.88 & 0.01 \\
SPT-CLJ2022-6323 &  305.5261 & $-63.3989$ & 305.53709 & $-63.39950$ & 0.383 & 6.17 & 4.85 & 0.01 \\
SPT-CLJ2043-5035 &  310.8284 & $-50.5938$ & 310.82312 & $-50.59232^a$ & 0.723 & 6.88 & 3.18 & 0.01 \\
SPT-CLJ2055-5456 &  313.9957 & $-54.9368$ & 313.98389 & $-54.92727$ & 0.139 & 6.99 & 11.80 & 0.01 \\
SPT-CLJ2130-6458 &  322.7280 & $-64.9767$ & 322.73421 & $-64.97787$ & 0.316 & 7.25 & 5.96 & 0.02 \\
SPT-CLJ2134-4238 &  323.5020 & $-42.6438$ & 323.50290 & $-42.64810$ & 0.196 & 8.85 & 9.45 & 0.02 \\
SPT-CLJ2148-6116 &  327.1812 & $-61.2780$ & 327.16178 & $-61.26553^b$ & 0.571 & 6.70 & 3.71 & 0.02 \\
SPT-CLJ2222-4834 &  335.7122 & $-48.5735$ & 335.71118 & $-48.57642^a$ & 0.652 & 8.22 & 3.62 & 0.01 \\
SPT-CLJ2232-5959 &  338.1487 & $-59.9903$ & 338.14086 & $-59.99810^a$ & 0.594 & 8.39 & 3.89 & 0.00 \\
SPT-CLJ2233-5339 &  338.3295 & $-53.6502$ & 338.31500 & $-53.65259^a$ & 0.440 & 8.23 & 4.81 & 0.04 \\
SPT-CLJ2259-6057 &  344.7528 & $-60.9546$ & 344.75411 & $-60.95951^a$ & 0.75 & 8.57 & 3.34 & 0.01 \\
SPT-CLJ2331-5051 &  352.9608 & $-50.8639$ & 352.96306 & $-50.86496^{ac}$ & 0.576 & 8.46 & 3.99 & 0.00 \\
SPT-CLJ2332-5358 &  353.1057 & $-53.9675$ & 353.11446 & $-53.97442$ & 0.402 & 7.89 & 5.08 & 0.01 \\
SPT-CLJ2352-4657 &  358.0631 & $-46.9569$ & 358.06778 & $-46.96021^a$ & 0.73 & 6.71 & 3.14 & 0.00 \\
		\hline
	\end{tabular}
    \begin{tablenotes}
      \small
      \item {\bf Notes.} $^a$Same BCG candidate visually selected in \cite{mcdonald16}. $^b$Different BCG candidate selected in \cite{mcdonald16} due to the selection procedure (See~\S~\ref{sec:bcgselection} for details). $^c$Cluster in pre-merging event.
    \end{tablenotes}
    \end{threeparttable}
\end{table*}

\begin{table*}
\scriptsize
	\centering
	\caption{Results from the Schechter luminosity function fitting using all cluster galaxies/red sequence alone, within a 0.5/1 R$_{200}$ projected radius, and centered in the BCG or the SZ centroid. Also included are the fits for all the clusters excluding the disturbed and relaxed sample (``Intermediate")}
	\label{tab:results}
	\begin{tabular}{lccccccccc} 
    \hline \hline 
               & \multicolumn{2}{c}{$z\leq 0.55$} &  \multicolumn{2}{c}{$z>0.55$}  & \multicolumn{2}{c}{all redshifts} &  &  & \\
		\hline
		Type &  $\alpha$ & $m^*$ & $\alpha$ & $m^*$ & $\alpha$ & $m^*$ & R$_{200}$ & Pop. & Center \\
		\hline
\hline
All & $-1.08^{+0.03}_{-0.02}$ & $-0.11^{+0.08}_{-0.07}$ & $-0.90^{+0.06}_{-0.05}$ & $0.08^{+0.11}_{-0.09}$ & $-1.03^{+0.02}_{-0.02}$ & $-0.08^{+0.06}_{-0.04}$ & 1 & all & BCG\\
Intermediate & $-1.06^{+0.03}_{-0.03}$ & $-0.11^{+0.10}_{-0.08}$ & $-0.98^{+0.06}_{-0.05}$ & $0.04^{+0.12}_{-0.12}$ & $-1.05^{+0.03}_{-0.02}$ & $-0.11^{+0.08}_{-0.06}$ & 1 & all & BCG\\
Unrelaxed & $-1.08^{+0.07}_{-0.07}$ & $0.20^{+0.19}_{-0.19}$ & $-1.07^{+0.15}_{-0.13}$ & $-0.06^{+0.37}_{-0.40}$ & $-1.07^{+0.06}_{-0.06}$ & $0.04^{+0.16}_{-0.16}$ & 1 & all & BCG\\
Relaxed & $-1.03^{+0.08}_{-0.08}$ & $-0.06^{+0.21}_{-0.21}$ & $-0.62^{+0.11}_{-0.11}$ & $0.36^{+0.16}_{-0.17}$ & $-0.94^{+0.07}_{-0.05}$ & $-0.05^{+0.13}_{-0.13}$ & 1 & all & BCG\\
Not Gaussian & $-1.09^{+0.20}_{-0.18}$ & $-0.49^{+0.58}_{-0.80}$ & $-0.90^{+0.24}_{-0.19}$ & $-0.11^{+0.45}_{-0.47}$ & $-1.00^{+0.15}_{-0.13}$ & $-0.30^{+0.36}_{-0.37}$ & 1 & all & BCG\\
Gaussian & $-0.99^{+0.13}_{-0.14}$ & $0.09^{+0.31}_{-0.38}$ & $-0.61^{+0.12}_{-0.11}$ & $0.52^{+0.17}_{-0.17}$ & $-0.73^{+0.09}_{-0.08}$ & $0.39^{+0.15}_{-0.14}$ & 1 & all & BCG\\
\hline
All & $-0.85^{+0.04}_{-0.02}$ & $0.22^{+0.05}_{-0.05}$ & $-0.68^{+0.04}_{-0.04}$ & $0.32^{+0.06}_{-0.06}$ & $-0.81^{+0.02}_{-0.01}$ & $0.24^{+0.04}_{-0.02}$ & 1 & red & BCG\\
Intermediate & $-0.84^{+0.03}_{-0.03}$ & $0.22^{+0.06}_{-0.05}$ & $-0.75^{+0.05}_{-0.06}$ & $0.21^{+0.08}_{-0.08}$ & $-0.82^{+0.03}_{-0.03}$ & $0.20^{+0.05}_{-0.04}$ & 1 & red & BCG\\
Unrelaxed & $-0.85^{+0.06}_{-0.07}$ & $0.18^{+0.13}_{-0.13}$ & $-0.61^{+0.13}_{-0.13}$ & $0.52^{+0.18}_{-0.19}$ & $-0.78^{+0.07}_{-0.06}$ & $0.34^{+0.12}_{-0.10}$ & 1 & red & BCG\\
Relaxed & $-0.78^{+0.08}_{-0.07}$ & $0.44^{+0.14}_{-0.14}$ & $-0.55^{+0.09}_{-0.09}$ & $0.48^{+0.12}_{-0.12}$ & $-0.70^{+0.06}_{-0.05}$ & $0.34^{+0.09}_{-0.09}$ & 1 & red & BCG\\
Not Gaussian & $-1.26^{+0.18}_{-0.16}$ & $-0.74^{+0.49}_{-0.78}$ & $-0.34^{+0.23}_{-0.21}$ & $0.82^{+0.22}_{-0.23}$ & $-0.66^{+0.15}_{-0.14}$ & $0.44^{+0.20}_{-0.21}$ & 1 & red & BCG\\
Gaussian & $-0.82^{+0.13}_{-0.13}$ & $0.38^{+0.23}_{-0.25}$ & $-0.52^{+0.09}_{-0.08}$ & $0.51^{+0.11}_{-0.10}$ & $-0.62^{+0.07}_{-0.05}$ & $0.42^{+0.10}_{-0.08}$ & 1 & red & BCG\\
\hline
All & $-1.02^{+0.03}_{-0.03}$ & $0.00^{+0.07}_{-0.06}$ & $-0.95^{+0.04}_{-0.04}$ & $-0.01^{+0.10}_{-0.09}$ & $-1.01^{+0.03}_{-0.02}$ & $-0.01^{+0.06}_{-0.04}$ & 1 & all & SZ\\
Intermediate & $-0.99^{+0.03}_{-0.03}$ & $0.05^{+0.09}_{-0.08}$ & $-0.98^{+0.07}_{-0.06}$ & $-0.06^{+0.14}_{-0.13}$ & $-1.01^{+0.03}_{-0.03}$ & $-0.06^{+0.07}_{-0.06}$ & 1 & all & SZ\\
Unrelaxed & $-1.11^{+0.09}_{-0.07}$ & $0.03^{+0.18}_{-0.19}$ & $-1.17^{+0.10}_{-0.09}$ & $-0.54^{+0.31}_{-0.35}$ & $-1.14^{+0.06}_{-0.05}$ & $-0.16^{+0.16}_{-0.17}$ & 1 & all & SZ\\
Relaxed & $-1.01^{+0.08}_{-0.08}$ & $0.04^{+0.19}_{-0.20}$ & $-0.71^{+0.12}_{-0.11}$ & $0.41^{+0.18}_{-0.19}$ & $-0.93^{+0.07}_{-0.06}$ & $0.08^{+0.14}_{-0.13}$ & 1 & all & SZ\\
Not Gaussian & $-1.03^{+0.21}_{-0.19}$ & $-0.03^{+0.47}_{-0.59}$ & $-0.92^{+0.22}_{-0.19}$ & $-0.16^{+0.46}_{-0.47}$ & $-0.98^{+0.15}_{-0.13}$ & $-0.14^{+0.33}_{-0.36}$ & 1 & all & SZ\\
Gaussian & $-0.85^{+0.14}_{-0.13}$ & $0.28^{+0.28}_{-0.32}$ & $-0.69^{+0.10}_{-0.11}$ & $0.44^{+0.16}_{-0.18}$ & $-0.75^{+0.09}_{-0.07}$ & $0.38^{+0.15}_{-0.13}$ & 1 & all & SZ\\
\hline
All & $-0.82^{+0.03}_{-0.03}$ & $0.25^{+0.05}_{-0.04}$ & $-0.67^{+0.04}_{-0.04}$ & $0.31^{+0.06}_{-0.06}$ & $-0.79^{+0.01}_{-0.02}$ & $0.24^{+0.03}_{-0.02}$ & 1 & red & SZ\\
Intermediate & $-0.80^{+0.04}_{-0.02}$ & $0.28^{+0.06}_{-0.05}$ & $-0.71^{+0.05}_{-0.06}$ & $0.24^{+0.09}_{-0.08}$ & $-0.80^{+0.03}_{-0.03}$ & $0.24^{+0.05}_{-0.04}$ & 1 & red & SZ\\
Unrelaxed & $-0.87^{+0.06}_{-0.06}$ & $0.16^{+0.12}_{-0.12}$ & $-0.80^{+0.10}_{-0.10}$ & $0.22^{+0.18}_{-0.18}$ & $-0.85^{+0.06}_{-0.05}$ & $0.20^{+0.11}_{-0.10}$ & 1 & red & SZ\\
Relaxed & $-0.77^{+0.08}_{-0.08}$ & $0.41^{+0.14}_{-0.14}$ & $-0.55^{+0.09}_{-0.08}$ & $0.51^{+0.12}_{-0.12}$ & $-0.69^{+0.05}_{-0.06}$ & $0.35^{+0.09}_{-0.09}$ & 1 & red & SZ\\
Not Gaussian & $-1.31^{+0.17}_{-0.15}$ & $-0.85^{+0.55}_{-0.85}$ & $-0.51^{+0.22}_{-0.20}$ & $0.59^{+0.24}_{-0.26}$ & $-0.79^{+0.14}_{-0.13}$ & $0.21^{+0.22}_{-0.24}$ & 1 & red & SZ\\
Gaussian & $-0.87^{+0.12}_{-0.10}$ & $0.30^{+0.23}_{-0.22}$ & $-0.57^{+0.08}_{-0.07}$ & $0.42^{+0.11}_{-0.09}$ & $-0.65^{+0.06}_{-0.06}$ & $0.36^{+0.09}_{-0.09}$ & 1 & red & SZ\\
\hline
All & $-0.97^{+0.03}_{-0.01}$ & $0.11^{+0.07}_{-0.04}$ & $-0.80^{+0.05}_{-0.05}$ & $0.26^{+0.09}_{-0.08}$ & $-0.91^{+0.02}_{-0.01}$ & $0.15^{+0.05}_{-0.02}$ & $\sfrac{1}{2}$ & all & BCG\\
Intermediate & $-0.94^{+0.03}_{-0.02}$ & $0.15^{+0.08}_{-0.04}$ & $-0.80^{+0.07}_{-0.06}$ & $0.34^{+0.11}_{-0.10}$ & $-0.93^{+0.02}_{-0.02}$ & $0.14^{+0.06}_{-0.03}$ & $\sfrac{1}{2}$ & all & BCG\\
Unrelaxed & $-0.98^{+0.12}_{-0.10}$ & $0.25^{+0.26}_{-0.28}$ & $-1.10^{+0.13}_{-0.11}$ & $-0.16^{+0.33}_{-0.34}$ & $-1.00^{+0.09}_{-0.09}$ & $0.15^{+0.20}_{-0.21}$ & $\sfrac{1}{2}$ & all & BCG\\
Relaxed & $-1.10^{+0.08}_{-0.08}$ & $-0.18^{+0.22}_{-0.25}$ & $-0.54^{+0.10}_{-0.11}$ & $0.60^{+0.13}_{-0.14}$ & $-0.83^{+0.06}_{-0.06}$ & $0.23^{+0.12}_{-0.11}$ & $\sfrac{1}{2}$ & all & BCG\\
Not Gaussian & $-1.70^{+0.23}_{-0.10}$ & $-8.70^{+7.49}_{-1.49}$ & $-0.94^{+0.25}_{-0.23}$ & $0.23^{+0.37}_{-0.47}$ & $-0.96^{+0.19}_{-0.17}$ & $0.10^{+0.34}_{-0.40}$ & $\sfrac{1}{2}$ & all & BCG\\
Gaussian & $-0.97^{+0.12}_{-0.12}$ & $-0.04^{+0.32}_{-0.33}$ & $-0.58^{+0.11}_{-0.10}$ & $0.52^{+0.16}_{-0.14}$ & $-0.70^{+0.09}_{-0.07}$ & $0.37^{+0.13}_{-0.13}$ & $\sfrac{1}{2}$ & all & BCG\\
\hline
All & $-0.80^{+0.03}_{-0.02}$ & $0.34^{+0.06}_{-0.03}$ & $-0.69^{+0.04}_{-0.04}$ & $0.31^{+0.07}_{-0.07}$ & $-0.77^{+0.02}_{-0.02}$ & $0.31^{+0.05}_{-0.03}$ & $\sfrac{1}{2}$ & red & BCG\\
Intermediate & $-0.78^{+0.04}_{-0.02}$ & $0.36^{+0.06}_{-0.04}$ & $-0.73^{+0.06}_{-0.06}$ & $0.15^{+0.09}_{-0.10}$ & $-0.78^{+0.02}_{-0.03}$ & $0.28^{+0.05}_{-0.03}$ & $\sfrac{1}{2}$ & red & BCG\\
Unrelaxed & $-0.82^{+0.12}_{-0.12}$ & $0.52^{+0.22}_{-0.23}$ & $-0.74^{+0.15}_{-0.14}$ & $0.48^{+0.23}_{-0.25}$ & $-0.77^{+0.09}_{-0.09}$ & $0.50^{+0.15}_{-0.15}$ & $\sfrac{1}{2}$ & red & BCG\\
Relaxed & $-0.81^{+0.08}_{-0.07}$ & $0.31^{+0.16}_{-0.16}$ & $-0.45^{+0.10}_{-0.10}$ & $0.69^{+0.13}_{-0.12}$ & $-0.64^{+0.06}_{-0.06}$ & $0.50^{+0.09}_{-0.09}$ & $\sfrac{1}{2}$ & red & BCG\\
Not Gaussian & $-1.28^{+0.20}_{-0.19}$ & $-0.93^{+0.64}_{-1.24}$ & $-0.28^{+0.29}_{-0.27}$ & $1.00^{+0.22}_{-0.25}$ & $-0.65^{+0.18}_{-0.18}$ & $0.62^{+0.23}_{-0.24}$ & $\sfrac{1}{2}$ & red & BCG\\
Gaussian & $-0.78^{+0.14}_{-0.13}$ & $0.37^{+0.25}_{-0.26}$ & $-0.49^{+0.10}_{-0.09}$ & $0.51^{+0.13}_{-0.12}$ & $-0.54^{+0.08}_{-0.07}$ & $0.46^{+0.10}_{-0.10}$ & $\sfrac{1}{2}$ & red & BCG\\
\hline
All & $-0.94^{+0.03}_{-0.01}$ & $0.18^{+0.07}_{-0.03}$ & $-0.84^{+0.05}_{-0.05}$ & $0.25^{+0.09}_{-0.09}$ & $-0.91^{+0.02}_{-0.01}$ & $0.18^{+0.04}_{-0.02}$ & $\sfrac{1}{2}$ & all & SZ\\
Intermediate & $-0.95^{+0.04}_{-0.01}$ & $0.13^{+0.07}_{-0.04}$ & $-0.91^{+0.06}_{-0.06}$ & $0.18^{+0.12}_{-0.12}$ & $-0.95^{+0.02}_{-0.02}$ & $0.11^{+0.06}_{-0.03}$ & $\sfrac{1}{2}$ & all & SZ\\
Unrelaxed & $-0.75^{+0.08}_{-0.08}$ & $0.65^{+0.15}_{-0.14}$ & $-0.73^{+0.14}_{-0.13}$ & $0.48^{+0.22}_{-0.25}$ & $-0.77^{+0.08}_{-0.08}$ & $0.60^{+0.14}_{-0.13}$ & $\sfrac{1}{2}$ & all & SZ\\
Relaxed & $-1.03^{+0.08}_{-0.08}$ & $-0.13^{+0.20}_{-0.20}$ & $-0.69^{+0.10}_{-0.09}$ & $0.45^{+0.15}_{-0.15}$ & $-0.85^{+0.06}_{-0.06}$ & $0.22^{+0.11}_{-0.11}$ & $\sfrac{1}{2}$ & all & SZ\\
Not Gaussian & $-1.32^{+0.23}_{-0.21}$ & $-0.73^{+0.56}_{-0.86}$ & $-0.84^{+0.27}_{-0.25}$ & $0.27^{+0.39}_{-0.53}$ & $-0.80^{+0.18}_{-0.15}$ & $0.24^{+0.27}_{-0.29}$ & $\sfrac{1}{2}$ & all & SZ\\
Gaussian & $-0.97^{+0.12}_{-0.10}$ & $-0.03^{+0.32}_{-0.31}$ & $-0.54^{+0.12}_{-0.11}$ & $0.75^{+0.15}_{-0.15}$ & $-0.64^{+0.09}_{-0.08}$ & $0.59^{+0.14}_{-0.13}$ & $\sfrac{1}{2}$ & all & SZ\\
\hline
All & $-0.79^{+0.02}_{-0.02}$ & $0.36^{+0.05}_{-0.03}$ & $-0.62^{+0.05}_{-0.04}$ & $0.43^{+0.08}_{-0.06}$ & $-0.71^{+0.01}_{-0.02}$ & $0.42^{+0.03}_{-0.02}$ & $\sfrac{1}{2}$ & red & SZ\\
Intermediate & $-0.78^{+0.03}_{-0.03}$ & $0.36^{+0.06}_{-0.04}$ & $-0.70^{+0.06}_{-0.06}$ & $0.25^{+0.10}_{-0.09}$ & $-0.78^{+0.03}_{-0.01}$ & $0.31^{+0.05}_{-0.03}$ & $\sfrac{1}{2}$ & red & SZ\\
Unrelaxed & $-0.76^{+0.07}_{-0.06}$ & $0.42^{+0.12}_{-0.12}$ & $-0.43^{+0.14}_{-0.13}$ & $0.84^{+0.16}_{-0.16}$ & $-0.63^{+0.08}_{-0.06}$ & $0.68^{+0.11}_{-0.09}$ & $\sfrac{1}{2}$ & red & SZ\\
Relaxed & $-0.76^{+0.08}_{-0.08}$ & $0.35^{+0.16}_{-0.15}$ & $-0.36^{+0.11}_{-0.09}$ & $0.86^{+0.12}_{-0.11}$ & $-0.62^{+0.06}_{-0.06}$ & $0.57^{+0.10}_{-0.09}$ & $\sfrac{1}{2}$ & red & SZ\\
Not Gaussian & $-1.07^{+0.23}_{-0.21}$ & $-0.30^{+0.53}_{-0.76}$ & $-0.39^{+0.27}_{-0.25}$ & $0.94^{+0.23}_{-0.25}$ & $-0.58^{+0.19}_{-0.18}$ & $0.78^{+0.21}_{-0.24}$ & $\sfrac{1}{2}$ & red & SZ\\
Gaussian & $-0.78^{+0.13}_{-0.11}$ & $0.43^{+0.22}_{-0.22}$ & $-0.44^{+0.10}_{-0.09}$ & $0.59^{+0.12}_{-0.12}$ & $-0.53^{+0.07}_{-0.07}$ & $0.48^{+0.10}_{-0.09}$ & $\sfrac{1}{2}$ & red & SZ\\
\hline
		\hline
	\end{tabular}
\end{table*}

\section*{Acknowledgements}
TJ is pleased to acknowledge funding support from DE-SC0010107. AS is supported by the ERC-StG `ClustersXCosmo' grant agreement 716762, and by the FARE-MIUR grant `ClustersXEuclid' R165SBKTMA.

Funding for the DES Projects has been provided by the U.S. Department of Energy, the U.S. National Science Foundation, the Ministry of Science and Education of Spain, the Science and Technology Facilities Council of the United Kingdom, the Higher Education Funding Council for England, the National Center for Supercomputing Applications at the University of Illinois at Urbana-Champaign, the Kavli Institute of Cosmological Physics at the University of Chicago, the Center for Cosmology and Astro-Particle Physics at the Ohio State University, the Mitchell Institute for Fundamental Physics and Astronomy at Texas A\&M University, Financiadora de Estudos e Projetos, Funda\c c\~ao Carlos Chagas Filho de Amparo \`a Pesquisa do Estado do Rio de Janeiro, Conselho Nacional de Desenvolvimento Cient\'ifico e Tecnol\'ogico and the Minist\'erio da Ci\^encia, Tecnologia e Inova\c c\~ao, the Deutsche Forschungsgemeinschaft and the Collaborating Institutions in the Dark Energy Survey.
The Collaborating Institutions are Argonne National Laboratory, the University of California at Santa Cruz, the University of Cambridge, Centro de Investigaciones Energ\'eticas, Medioambientales y Tecnol\'ogicas-Madrid, the University of Chicago, University College London, the DES-Brazil Consortium, the University of Edinburgh, the Eidgen\"ossische Technische Hochschule (ETH) Z\"urich, Fermi National Accelerator Laboratory, the University of Illinois at Urbana-Champaign, the Institut de Ci\'encies de l'Espai (IEEC/CSIC), the Institut de F\'isica d'Altes Energies, Lawrence Berkeley National Laboratory, the Ludwig-Maximilians Universit\"at M\"unchen and the associated Excellence Cluster Universe, the University of Michigan, the National Optical Astronomy Observatory, the University of Nottingham, The Ohio State University, the University of Pennsylvania, the University of Portsmouth, SLAC National Accelerator Laboratory, Stanford University, the University of Sussex, Texas A\&M University, and the OzDES Member- ship Consortium.
Based in part on observations at Cerro Tololo Inter-American Observatory, National Optical Astronomy Observatory, which is operated by the Association of Universities for Research in Astronomy (AURA) under a cooperative agreement with the National Science Foundation.
The DES data management system is supported by the National Science Foundation under Grant Numbers AST-1138766 and AST-1536171. The DES participants from Spanish institutions are partially supported by MINECO under grants AYA2015- 71825, ESP2015-66861, FPA2015-68048, SEV-2016-0588, SEV- 2016-0597, and MDM-2015-0509, some of which include ERDF funds from the European Union. IFAE is partially funded by the CERCA program of the Generalitat de Catalunya. Research leading to these results has received funding from the European Research Council under the European Union’s Seventh Framework Program (FP7/2007-2013) including ERC grant agreements 240672, 291329, and 306478. We acknowledge support from the Australian Research Council Centre of Excellence for All-sky Astrophysics (CAASTRO), through project number CE110001020.
This manuscript has been authored by Fermi Research Alliance, LLC under Contract No. DE-AC02-07CH11359 with the U.S. Department of Energy, Office of Science, Office of High Energy Physics. The United States Government retains and the publisher, by accepting the article for publication, acknowledges that the United States Government retains a non-exclusive, paid-up, irrevocable, world-wide license to publish or reproduce the published form of this manuscript, or allow others to do so, for United States Government purposes.



\bibliographystyle{mn2e_2author_amp.bst}
\bibliography{mnras_template} 




\clearpage
\parbox{\textwidth}{
\scriptsize
$^{18}$ Departamento de F\'isica Matem\'atica, Instituto de F\'isica, Universidade de S\~ao Paulo, CP 66318, S\~ao Paulo, SP, 05314-970, Brazil\\
$^{19}$ Laborat\'orio Interinstitucional de e-Astronomia - LIneA, Rua Gal. Jos\'e Cristino 77, Rio de Janeiro, RJ - 20921-400, Brazil\\
$^{20}$ Instituto de Fisica Teorica UAM/CSIC, Universidad Autonoma de Madrid, 28049 Madrid, Spain\\
$^{21}$ Department of Physics, University of Cincinnati, Cincinnati, OH 45221, USA\\
$^{22}$ CNRS, UMR 7095, Institut d'Astrophysique de Paris, F-75014, Paris, France\\
$^{23}$ Sorbonne Universit\'es, UPMC Univ Paris 06, UMR 7095, Institut d'Astrophysique de Paris, F-75014, Paris, France\\
$^{24}$ Department of Physics \& Astronomy, University College London, Gower Street, London, WC1E 6BT, UK\\
$^{25}$ Centro de Investigaciones Energ\'eticas, Medioambientales y Tecnol\'ogicas (CIEMAT), Madrid, Spain\\
$^{26}$ Department of Astronomy, University of Illinois at Urbana-Champaign, 1002 W. Green Street, Urbana, IL 61801, USA\\
$^{27}$ National Center for Supercomputing Applications, 1205 West Clark St., Urbana, IL 61801, USA\\
$^{28}$ Institut de F\'{\i}sica d'Altes Energies (IFAE), The Barcelona Institute of Science and Technology, Campus UAB, 08193 Bellaterra (Barcelona) Spain\\
$^{29}$ Institut d'Estudis Espacials de Catalunya (IEEC), 08034 Barcelona, Spain\\
$^{30}$ Institute of Space Sciences (ICE, CSIC),  Campus UAB, Carrer de Can Magrans, s/n,  08193 Barcelona, Spain\\
$^{31}$ Institute for Fundamental Physics of the Universe, Via Beirut 2, 34014 Trieste, Italy\\
$^{32}$ Observat\'orio Nacional, Rua Gal. Jos\'e Cristino 77, Rio de Janeiro, RJ - 20921-400, Brazil\\
$^{33}$ Department of Physics, IIT Hyderabad, Kandi, Telangana 502285, India\\
$^{34}$ Department of Astronomy/Steward Observatory, University of Arizona, 933 North Cherry Avenue, Tucson, AZ 85721-0065, USA\\
$^{35}$ Jet Propulsion Laboratory, California Institute of Technology, 4800 Oak Grove Dr., Pasadena, CA 91109, USA\\
$^{36}$ Department of Astronomy, University of Michigan, Ann Arbor, MI 48109, USA\\
$^{37}$ Department of Physics, University of Michigan, Ann Arbor, MI 48109, USA\\
$^{38}$ Department of Physics and Astronomy, University of Missouri, Kansas City\\
$^{39}$ Department of Astronomy, University of Michigan, 1085 S. University, Ann Arbor, MI 48109, USA\\
$^{40}$ D\'{e}partement de Physique Th\'{e}orique and Center for Astroparticle Physics, Universit\'{e} de Gen\`{e}ve, 24 quai Ernest Ansermet, CH-1211 Geneva, Switzerland\\
$^{41}$ Department of Physics, ETH Zurich, Wolfgang-Pauli-Strasse 16, CH-8093 Zurich, Switzerland\\
$^{42}$ School of Mathematics and Physics, University of Queensland,  Brisbane, QLD 4072, Australia\\
$^{43}$ Center for Cosmology and Astro-Particle Physics, The Ohio State University, Columbus, OH 43210, USA\\
$^{44}$ Department of Physics, The Ohio State University, Columbus, OH 43210, USA\\
$^{45}$ Center for Astrophysics $\vert$ Harvard \& Smithsonian, 60 Garden Street, Cambridge, MA 02138, USA\\
$^{46}$ Australian Astronomical Optics, Macquarie University, North Ryde, NSW 2113, Australia\\
$^{47}$ Lowell Observatory, 1400 Mars Hill Rd, Flagstaff, AZ 86001, USA\\
$^{48}$ Kavli Institute for Astrophysics and Space Research, Massachusetts Institute of Technology, 77 Massachusetts Avenue, Cambridge, MA 02139, USA\\
$^{49}$ Department of Physics and Astronomy, University of Pennsylvania, Philadelphia, PA 19104, USA\\
$^{50}$ Department of Astrophysical Sciences, Princeton University, Peyton Hall, Princeton, NJ 08544, USA\\
$^{51}$ Instituci\'o Catalana de Recerca i Estudis Avan\c{c}ats, E-08010 Barcelona, Spain\\
$^{52}$ Institute of Astronomy, University of Cambridge, Madingley Road, Cambridge CB3 0HA, UK\\
$^{53}$ School of Physics and Astronomy, University of Southampton,  Southampton, SO17 1BJ, UK\\
$^{54}$ Brandeis University, Physics Department, 415 South Street, Waltham MA 02453\\
$^{55}$ Computer Science and Mathematics Division, Oak Ridge National Laboratory, Oak Ridge, TN 37831\\
$^{56}$ Institute of Cosmology and Gravitation, University of Portsmouth, Portsmouth, PO1 3FX, UK\\
}






\bsp	
\label{lastpage}
\end{document}